\documentclass[12pt,showpacs,showkeys,amsmath,amssymb]{revtex4}
\usepackage{amsmath,amsfonts,amsthm,amscd,amssymb,latexsym}
\usepackage{bm}
\usepackage{dcolumn}
\usepackage{graphicx}
\usepackage{epstopdf}
\usepackage{color}
\usepackage{epsf}
\usepackage{epsfig}
\usepackage{graphicx, epic, eepic, color}
\usepackage[colorlinks,citecolor=blue,urlcolor=blue,linkcolor=blue]{hyperref}

\newcommand{\beq}{\begin{equation}}
\newcommand{\eeq}{\end{equation}}

\begin{document}

\title{Spin-Gravity Coupling in a Rotating Universe}

\author{Bahram \surname{Mashhoon}$^{1,2}$}
\email{mashhoonb@missouri.edu}
\author{Masoud \surname{Molaei}$^{3}$}
\email{masoud.molaei@sharif.ir}
\author{Yuri N. Obukhov$^4$}
\email{obukhov@ibrae.ac.ru}

\affiliation{$^1$Department of Physics and Astronomy, University of Missouri, Columbia, Missouri 65211, USA\\
$^2$School of Astronomy, Institute for Research in Fundamental
Sciences (IPM), Tehran 19395-5531, Iran\\
$^3$Department of Physics, Sharif University of Technology, Tehran 11365-9161, Iran\\
$^4$Theoretical Physics Laboratory, Nuclear Safety Institute, \\
Russian Academy of Sciences, B. Tulskaya 52, 115191 Moscow, Russia
}

\date{\today}

\begin{abstract}
The coupling of intrinsic spin with the nonlinear gravitomagnetic fields of G\"odel-type spacetimes is studied.  We work with G\"odel-type universes in order to show that the main features of spin-gravity coupling are independent of causality problems of the G\"odel universe.  The connection between the  spin-gravitomagnetic field coupling and Mathisson's spin-curvature force is demonstrated in the G\"odel-type universe. That is, the gravitomagnetic Stern--Gerlach force due to the coupling of spin with the gravitomagnetic field reduces in the appropriate correspondence limit to the classical Mathisson spin-curvature force. 
\end{abstract}


\maketitle

\section{Introduction}

Inertia is the intrinsic tendency of matter to remain in a given condition. The state of matter in spacetime is determined by its mass and spin; indeed, mass and spin characterize the irreducible unitary representations of the Poincar\'e group~\cite{Wigner}. Therefore, mass and spin determine the inertial properties of a particle. In classical physics, the inertial forces that act on a particle are proportional to its inertial mass; moreover, the moment of inertia is the rotational analogue of mass. Inertial effects of intrinsic spin are independent of the inertial mass of the particle and depend purely on intrinsic spin. Inertia of intrinsic spin is of quantum origin and its properties therefore complement the inertial characteristics of mass and orbital angular momentum of the particle. 

It turns out that the intrinsic spin $\bm{S}$ of a particle couples to the rotation of a noninertial observer resulting in a Hamiltonian of the form $\mathcal{H}_{sr} =  -\,\bm{S} \cdot \bm{\Omega}$, where $\bm{\Omega}$ is the angular velocity of the observer's local spatial frame with respect to a nonrotating (i.e., Fermi--Walker) transported frame. For an intuitive explanation of this type of coupling, let us consider a noninertial observer that is at rest in Minkowski spacetime but refers its observations to axes that rotate uniformly  with angular speed $\Omega$ in the positive sense about the direction of propagation of a plane electromagnetic wave of frequency $\omega > \Omega$. The Fourier analysis of the electromagnetic field detected by the noninertial observer reveals that the measured frequency of the wave is given by $\omega \mp \Omega$, where the upper (lower) sign refers to positive (negative) helicity radiation. One can understand this result as a kind of ``rotational Doppler effect": In a positive (negative) helicity electromagnetic wave, the electric and magnetic fields rotate in the positive (negative) sense with the wave frequency $\omega$ about the direction of propagation. The noninertial observer thus realizes that the positive (negative) helicity radiation has electric and magnetic fields that rotate in the positive (negative) sense with frequency $\omega - \Omega$ ($\omega + \Omega$) about the direction of wave propagation.  Multiplication of the measured frequency by $\hbar$ results in the measured energy by the noninertial observer, namely,  $\hbar \,\omega \mp \hbar \,\Omega$, which illustrates the coupling of photon helicity with rotation. A general  consequence of spin-rotation coupling should be noted here: There is a certain shift in energy when polarized radiation passes through a rotating spin flipper. To demonstrate this effect within the context of the present discussion, imagine the noninertial observer is replaced by a uniformly rotating half-wave plate. That is, electromagnetic radiation of frequency $\omega_{\rm in}$ is normally  incident on the plate rotating with $\Omega < \omega_{\rm in}$. The frequency of the radiation within the stationary medium of the half-wave plate remains constant and approximately equal to $\omega_{\rm in} \mp \Omega$, since we have neglected time dilation for simplicity. The outgoing radiation has opposite helicity to the incident radiation and frequency $\omega_{\rm out}$, where  $\omega_{\rm in} \mp \Omega \approx \omega_{\rm out} \pm \Omega$ due to helicity-rotation coupling. Therefore, $\omega_{\rm out} - \omega_{\rm in} \approx \mp \,2\, \Omega$ and the photon energy in passing through the rotating half-wave plate  is shifted by $ \approx \mp \,2\, \hbar\,\Omega$.  

A general account of the spin-rotation coupling is contained in~\cite{BMB} and  more recent discussions of its observational basis can be found in~\cite{DSH, DDSH, DDKWLSH, Yu:2022vjn}. A similar phenomenon occurs in a gravitational field~\cite{DeOT, HN, Soares:1995cj}. The spin-rotation effect can be theoretically extended to the spin-gravity coupling via the gravitational Larmor theorem~\cite{Larmor, Bahram}, which is the rotational side of Einstein's principle of equivalence. Imagine a free test gyroscope with its center of mass held at rest in a gravitational field; then, the locally measured components of the gyroscope's spin vector undergo a precessional motion with an angular velocity that is given by the locally measured gravitomagnetic field. The Gravity Probe B (GP-B) space experiment has measured the gravitomagnetic field of the Earth~\cite{Francis1, Francis2}. 

According to the gravitational Larmor theorem, the gravitomagnetic field of a rotating system is locally equivalent to a rotation resulting in a Hamiltonian for intrinsic spin-gravity coupling of the form $\mathcal{H}_{sg} = \bm{S}\cdot\bm{B}$, where $\bm{B}$ is the relevant gravitomagnetic field~\cite{Papini:2007gx}. The spin-gravity coupling is of basic physical significance due to the fundamental nature of intrinsic spin of particles and the universality of the gravitational interaction. For prospects regarding the measurement of intrinsic spin-gravity coupling, see~\cite{Bah1, Bah2, Tarallo:2014oaa, Fadeev:2020gjk, Vergeles:2022mqu}. In general, $\bm{B}$ depends on position and the intrinsic spin-gravity coupling leads to a measured gravitomagnetic Stern--Gerlach force of the form $- \bm{\nabla}(\bm{S} \cdot \bm{B})$. This gravitational force which acts on a test particle is completely independent of its inertial mass and depends solely on its intrinsic spin.  It has been shown~\cite{Mashhoon:2021qtc}, within the framework of  \emph{linearized} general relativity, that the gravitomagnetic Stern--Gerlach force associated with spin-gravity coupling reduces in the correspondence limit  to Mathisson's classical spin-curvature force~\cite{Math, Mashhoon:2008si}.   It would be interesting to extend this result to the nonlinear regime. The purpose of the present work is to study further the inertial effects of intrinsic spin by investigating the intrinsic spin-gravity coupling  for spinning test particles in G\"odel-type spacetimes. For background material,  Ref.~\cite{Mashhoon:2021qtc} and the references cited therein should be consulted for further important  information regarding the topic of spin-rotation-gravity coupling and its experimental basis. 


\section{Gravitomagnetism in the G\"odel-type Universe}

With respect to spacetime coordinates $x^\mu = (ct, x, y, z)$, the metric  of the G\"odel solution~\cite{Goedel} of Einstein's gravitational field equations arises as a special case in the class of the so-called G\"odel-type models \cite{RT:1983,RT:1985,YNO} described by the line element
\begin{equation}\label{GT} 
ds^2 = g_{\mu\nu}dx^\mu dx^\nu = -\,dt^2 -2\sqrt{\sigma}\,e^{\mu x}\,dt\,dy + dx^2 + \kappa\, e^{2\mu x}\,dy^2 +dz^2\,,
\end{equation}
with arbitrary constant parameters $\mu, \sigma$ and $\kappa$. In our conventions, the speed of light $c = 1$ and Planck's constant $\hbar = 1$, unless specified otherwise; moreover,  the metric signature is +2 and Greek indices run from 0 to 3, while Latin indices run from 1 to 3. The system of coordinates in metric~\eqref{GT} is admissible provided
\begin{equation}
\sigma + \kappa > 0\,.\label{ad}
\end{equation}
Moreover, we assume throughout that $\sigma > 0$. 
In general, the G\"odel-type universe contains closed timelike curves, which could lead to problems with causality. However, one can demonstrate \cite{YNO} that closed timelike curves are absent in model (\ref{GT}), provided
\begin{equation}
\kappa \geq 0\,.\label{CTC}
\end{equation}

Specifically, for the G\"odel universe $\kappa = -1$ in metric~\eqref{GT}; therefore, closed timelike curves do exist in the G\"odel universe. To ensure that our considerations regarding spin-gravity coupling are independent of the causality difficulties of the G\"odel universe, we use metric~\eqref{GT} for our main calculations in this paper. 

The G\"odel-type universe is a regular stationary and spatially homogeneous spacetime that contains rotating matter. Consider the class of observers that are all spatially at rest in this spacetime. Each such observer has a velocity 4-vector $u^\mu = \delta^\mu_0$ that is free of acceleration, expansion and shear; however, it is rotating in the negative sense about the $z$ axis and its vorticity 4-vector
\begin{equation}\label{G12} 
\omega^\mu  =  \frac{1}{2} \eta^{\mu \nu \rho \sigma}u_\nu  u_{\rho; \sigma}\,, 
\end{equation}
is purely spatial $\omega^\mu = (0, \bm{\omega})$, with the 3-vector 
\begin{equation}\label{G13} 
\bm{\omega}  = -\,\Omega\,\partial_z\,,\qquad \Omega = {\frac \mu 2}\sqrt{\frac {\sigma}{\sigma + \kappa}}\,. 
\end{equation}
For the sake of definiteness, we henceforth assume that $\Omega > 0$; then, Eq.~\eqref{G13} implies that $\mu > 0$ as well. Here, $\eta_{\alpha \beta \gamma \delta} = (-g)^{1/2} \epsilon_{\alpha \beta \gamma \delta}$ is the Levi-Civita tensor and  $\epsilon_{\alpha \beta \gamma \delta}$ is the alternating symbol with $\epsilon_{0123} = 1$.
It is interesting to note that in nonrelativistic fluid mechanics vorticity vector $\bm{\omega}_N$ is defined as $\bm{\omega}_N = \bm{\nabla} \times \bm{v}$, where $\bm{v}$ is the flow velocity. If the fluid rotates with spatially uniform angular velocity $\bm{\Omega}$  such that $\bm{v} = \bm{\Omega} \times \bm{x}$,  then $\bm{\omega}_N = 2 \,\bm{\Omega}$. In this paper, we follow the relativistic definition of vorticity. 

The geometry of the G\"odel-type model has been studied by a number of authors~\cite{HaEl,  Stephani:2003tm, GrPo}. The Weyl curvature of G\"odel-type spacetime is of type D in the Petrov classification. The G\"odel-type universe admits five Killing vector fields, namely, $\partial_t$, $\partial_y$, $\partial_z$, $\partial_x - \mu y\,\partial_y$ and~\cite{YNO, Chicone:2005vn} 
\begin{equation}\label{G7} 
K = {\frac {2\sqrt{\sigma}\,e^{-\,\mu x}}{\sigma + \kappa}}\,\partial_t - 2\mu y\,\partial_x
+ \left(\mu^2y^2 - {\frac {e^{-2\mu x}}{\sigma + \kappa}}\right)\partial_y\,.  
\end{equation}

We are interested in the measurements of an observer that is free and spatially at rest in spacetime with 4-velocity vector $u^\mu = dx^\mu/d\tau$ and proper time $\tau$, where $\tau = t + {\rm constant}$. The observer carries along its geodesic world line a natural tetrad frame $e^{\mu}{}_{\hat \alpha}$ that is orthonormal, namely, 
\begin{equation}\label{G8} 
g_{\mu \nu}\, e^{\mu}{}_{\hat \alpha} \,e^{\nu}{}_{\hat \beta} = \eta_{\hat \alpha \hat \beta}\,, 
\end{equation}
where $\eta_{\mu \nu} = \rm{diag}(-1, 1, 1, 1)$ is the Minkowski metric tensor. Indeed, 
\begin{equation}\label{G9} 
e_{\hat 0} = \partial_t\,, \qquad e_{\hat 1} = \partial_x\,, \qquad e_{\hat 2} =
-\,\sqrt{\frac {\sigma}{\sigma + \kappa}}\,\partial_t + {\frac {e^{-\,\mu x}}{\sqrt{\sigma + \kappa}}}
\,\partial_y\,,\qquad e_{\hat 3} = \partial_z\,,
\end{equation}
where the spatial axes of the observer's frame are primarily along the background coordinate axes. Introducing the dual coframe $\vartheta^{\hat\alpha}$,
\begin{equation}\label{cof}
\vartheta^{\widehat 0} = dt + \sqrt{\sigma}\,e^{\mu x}dy,\qquad \vartheta^{\widehat 1} = dx,\qquad
\vartheta^{\widehat 2} = \sqrt{\sigma + \kappa}\,e^{\mu x}d y,\qquad \vartheta^{\widehat 3} = dz,
\end{equation}
such that $e_{\hat\alpha}\rfloor\vartheta^{\hat\beta} = \delta_\alpha^\beta$, line element (\ref{GT}) is recast into
\begin{equation}\label{GT1}
ds^2 = -\,\left(dt + \sqrt{\sigma}\,e^{\mu x}\,dy\right)^2 + dx^2
+ (\sigma + \kappa)\,e^{2\mu x}\,dy^2 +dz^2\,.
\end{equation}

Let $\lambda^{\mu}{}_{\hat \alpha}$ be the orthonormal tetrad frame that is parallel transported along the observer's geodesic world line such that $D\lambda^{\mu}{}_{\hat \alpha}/d\tau = 0$. We find that 
\begin{equation}\label{G10} 
\lambda^{\mu}{}_{\hat 1}  =  e^{\mu}{}_{\hat 1} \cos \Omega \tau + e^{\mu}{}_{\hat 2} \sin \Omega \tau\,, \qquad \lambda^{\mu}{}_{\hat 2}  = - e^{\mu}{}_{\hat 1} \sin \Omega \tau + e^{\mu}{}_{\hat 2} \cos \Omega \tau\,,
\end{equation}
while  $\lambda^{\mu}{}_{\hat 3} = e^{\mu}{}_{\hat 3}$ and naturally $\lambda^{\mu}{}_{\hat 0} = e^{\mu}{}_{\hat 0} = u^\mu$. It is simple to check these results using the Christoffel symbols
\begin{equation}\label{G11} 
\Gamma^0_{10} = \sqrt{\frac {\sigma}{\sigma + \kappa}}\,\Omega\,,\qquad
\Gamma^1_{20} = \sqrt{\sigma + \kappa}\,e^{\mu x}\,\Omega\,, \qquad
\Gamma^2_{10} = -\,{\frac {e^{-\,\mu x}}{\sqrt{\sigma + \kappa}}}\,\Omega\,\,,  
\end{equation}
which are the only nonzero components of $\Gamma^\mu_{\nu 0}$. Therefore, the observer's natural frame rotates with respect to the parallel-transported frame about their common $z$ axis with frequency $-\Omega$, which is consistent with vorticity (\ref{G13}). 

Let us now consider the special case of metric~\eqref{GT} with parameters
\begin{equation}\label{msk}
\mu = \sqrt{2}\,\Omega,\qquad \sigma = 2,\qquad \kappa = -\,1\,.
\end{equation}
With these parameters, metric (\ref{GT}) reduces to the G\"odel line element
\begin{equation}\label{G1} 
ds^2 = -\,dt^2 -2\sqrt{2} \,e^{\sqrt{2} \Omega x}\,dt\,dy + dx^2 -e^{2\sqrt{2} \Omega x}\,dy^2 +dz^2\,. 
\end{equation}
For the G\"odel universe, Einstein's field equations
\begin{equation}\label{G3} 
R_{\mu \nu} - \frac{1}{2} g_{\mu \nu} R + \Lambda g_{\mu \nu} = 8 \pi G\,T_{\mu \nu}\, 
\end{equation}
have a perfect fluid source
\begin{equation}\label{G4} 
T_{\mu \nu} = (\rho + p) u_\mu u_\nu + p g_{\mu \nu}\,, 
\end{equation}
where $\rho$ is the energy density, $p$ is the pressure and $u^\mu = \delta^\mu_0$ is the 4-velocity vector of the perfect fluid. In this special case, $R_{\mu \nu} = 2 \Omega^2 u_\mu u_\nu$ and 
\begin{equation}\label{G6} 
2\, \Omega^2 = 8\pi G (\rho + p)\,, \qquad  \Lambda + \Omega^2 =  8\pi G  p\,.  
\end{equation}
In the absence of the cosmological constant $\Lambda$, we have as the source of  the G\"odel universe a perfect fluid with a stiff equation of state $\rho = p = \Omega^2/(8\pi G)$. Another possibility is dust ($p = 0$), with $4 \pi G \rho = - \Lambda = \Omega^2$. It follows from Eq.~\eqref{G6} that $-\Lambda = 4 \pi G (\rho - p)$; therefore, in any realistic situation, the cosmological constant of the G\"odel universe must be negative or zero ($\Lambda \le 0$).

The spinning test particle in the G\"odel universe is immersed in the perfect fluid source and its intrinsic spin couples to the vorticity of the fluid. The nature of the spin-gravity coupling and its connection with Mathisson's classical spin-curvature force provided the original motivation for the present work. 

After this brief digression regarding the G\"odel universe, we return to the G\"odel-type metric with explicit components 
\begin{equation}\label{G1a}
(g_{\mu \nu}) = 
\begin{bmatrix}
-1&0&-\sqrt{\sigma} \,W&0 \\
0&1&0&0 \\
-\sqrt{\sigma}\, W&0&\kappa\,W^2&0 \\
0&0&0&1 
\end{bmatrix}
\,, \quad (g^{\mu \nu}) =
\begin{bmatrix}
-\tfrac{\kappa}{\sigma + \kappa}&0&-\tfrac{\sqrt{\sigma}}{\sigma + \kappa} W^{-1}&0 \\
0&1&0&0 \\
-\tfrac{\sqrt{\sigma}}{\sigma + \kappa} W^{-1}&0&\tfrac{1}{\sigma + \kappa}W^{-2}&0 \\
0&0&0&1 
\end{bmatrix}\,,
\end{equation}
where $W(x) = e^{\mu x}$ and $\sqrt{-g} = \sqrt{\sigma + \kappa}\, W(x)$.  


\section{Mathisson's Spin-Curvature Force}

To connect Mathisson's classical spin-curvature force in the correspondence limit with intrinsic spin that is purely of quantum origin, it proves useful to introduce a classical model of intrinsic spin. To simplify matters, we permanently attach a free spin vector $\bm{S}$ to a Newtonian point particle resulting in a ``pole-dipole" particle. The particle thus carries the spin vector along its world line and the corresponding equations of motion in a gravitational field  are the  Mathisson--Papapetrou pole-dipole equations~\cite{Math, Papa},
\begin{equation}\label{M1} 
\frac{DP^\mu}{d\varsigma} =  -\,\frac{1}{2} R^{\mu}{}_{\nu \alpha \beta}\,U^\nu S^{\alpha \beta}\,,
\end{equation}
\begin{equation}\label{M2} 
\frac{DS^{\mu \nu}}{d\varsigma} = P^\mu U^\nu - P^\nu U^\mu\,,
\end{equation}
where  $U^\mu = dx^\mu/d\varsigma$ is the 4-velocity of the pole-dipole particle, $U^\mu U_\mu = -\,1$, and $\varsigma$ is its proper time. The particle's 4-momentum is $P^\mu$ and its spin tensor is $S^{\mu \nu}$, which satisfies the Frenkel--Pirani supplementary condition~\cite{Frenkel, Pirani} 
\begin{equation}\label{M3} 
S^{\mu \nu}\,U_\nu = 0\,.
\end{equation}
In this system, the inertial mass of the particle $m$, $m := -P^\mu U_\mu$, and the magnitude of its spin $s$, $s^2 := \tfrac{1}{2} S^{\mu \nu}S_{\mu \nu}$, are constants of the motion. Moreover, Pirani has shown that the spin vector $S^\mu$, $S^\mu\, U_\mu = 0$, 
\begin{equation}\label{M4} 
S^\mu = -\,\frac{1}{2} \,\eta^{\mu \nu \rho \sigma}\,U_\nu S_{\rho \sigma}\,,
\qquad S^{\alpha \beta} = \eta^{\alpha \beta \gamma \delta}\,U_\gamma S_\delta\,,
\end{equation}
is Fermi--Walker transported along the particle's world line~\cite{Pirani}. That is, the Mathisson--Papapetrou equations for a spinning test particle together with the Frenkel--Pirani supplementary condition imply that the spin vector of a test pole-dipole particle is nonrotating in this classical model consistent with the inertia of intrinsic spin. Furthermore,  the Mathisson--Papapetrou equations together with the Frenkel--Pirani supplementary condition imply that in the massless limit,  the spinning massless test  particle follows a null geodesic with the spin vector parallel or antiparallel to its direction of motion~\cite{Mash}.  Hence, our classical model is consistent with physical expectations. 

What is the influence of the inertia of intrinsic spin on the motion of the spinning particle? From Eq.~\eqref{M2}, we find
\begin{equation}\label{M5} 
P^\mu = m\,U^\mu + S^{\mu \nu}\,\frac{DU_\nu}{d\varsigma}\,,
\end{equation}
so that in the absence of spin, $P^\mu = m\,U^\mu$, and the particle simply follows a timelike geodesic of the background gravitational field. In the presence of spin, on the other hand, the Mathisson spin-curvature force $\mathcal{F}^\mu$, $\mathcal{F}^\mu U_\mu = 0$, 
\begin{equation}\label{M6} 
\mathcal{F}^\mu = -\,\frac{1}{2} R^{\mu}{}_{\nu \alpha \beta}\,U^\nu S^{\alpha \beta} =
~\,^*R^{\mu}{}_{\nu \rho \sigma}\,U^\nu \,S^{\rho}\,U^\sigma\,, \qquad ^*R_{\mu \nu \rho \sigma} =
\frac{1}{2}\,\eta_{\mu \nu \alpha \beta}\,R^{\alpha \beta}{}_{\rho \sigma}\,,
\end{equation}
must be taken into account~\cite{Mashhoon:2008si}. It follows from Eq.~\eqref{M5} that $P^\mu - m\,U^\mu$ is of second order in spin; hence,  the Mathisson--Papapetrou equations of motion to first order in spin become~\cite{CMP}
\begin{equation}\label{M7} 
\frac{DS^{\mu \nu}}{d\varsigma} \approx 0\, 
\end{equation} 
and
\begin{equation}\label{M8} 
m\,\frac{DU^\mu}{d\varsigma} \approx \mathcal{F}^\mu = -\,\frac{1}{2} R^{\mu}{}_{\nu \alpha \beta}\,U^\nu S^{\alpha \beta}\,. 
\end{equation}
 

\section{Spin-Vorticity-Gravity Coupling}\label{SVG}

We now turn to the behavior of spinning test  particles in the G\"odel-type spacetime. Within the framework of linearized general relativity, it can be shown in general that in source-free Ricci-flat regions of the gravitational field the Mathisson force corresponds to the Stern--Gerlach force associated with the spin-gravitomagnetic field coupling~\cite{Mashhoon:2021qtc}. In the G\"odel-type universe, on the other hand,  the spinning particle is immersed in the   source of the gravitational field. 
Is $\mathcal{F}_\mu = -\partial_\mu (\mathcal{H}_{sg})$ still valid for the G\"odel-type spacetime? 

Let us consider a spinning test particle held at rest in space at fixed $(x, y, z)$ coordinates in the G\"odel-type spacetime. According to the free reference observer with adapted tetrad frames $e^{\mu}{}_{\hat \alpha}$ and $\lambda^{\mu}{}_{\hat \alpha}$ at the same location, the spin vector to linear order stays fixed with respect to the parallel-propagated frame as a consequence of Eq.~\eqref{M7}; that is, $S_{\hat i}$, $i = 1, 2, 3$, are constants of the motion, where
$S_{\hat \alpha} = S_\mu\,\lambda^{\mu}{}_{\hat \alpha}$; hence, 
\begin{equation}\label{V1} 
 S_{\hat 0} = 0\,, \qquad S_{\hat i} = S_\mu\,\lambda^{\mu}{}_{\hat i}\,.
\end{equation}

The motion of the comoving observer has vorticity in accordance with Eq.~\eqref{G12} and we therefore expect that the spin should couple to the vorticity resulting in the spin-vorticity Hamiltonian given by 
\begin{equation}\label{V2}
\mathcal{H}_{sv} =  - \bm{S} \cdot\bm{\omega}   = \Omega \,S^{\hat 3}\,.
\end{equation}
Furthermore, the spin vector precesses with frequency $\Omega\, \partial_z$ with respect to the observer's natural frame $e^{\mu}{}_{\hat i}$ based on the spatial coordinate axes. The Hamiltonian associated with this motion is the spin-gravity Hamiltonian given by 
\begin{equation}\label{V3}
\mathcal{H}_{sg} =  \bm{S} \cdot \bm{B}\,,
\end{equation}
where $\bm{B} =\bm{\Omega} = \Omega\, \partial_z$ is the gravitomagnetic field in this case. The result is, 
\begin{equation}\label{V4}
\mathcal{H}_{sg} =   \Omega\,S^{\hat 3}\,.
\end{equation} 
The spin-gravity coupling is indeed the same as spin-vorticity coupling in this case, since the spinning particle, while engulfed by the source of the gravitational field, is fixed in space and comoving with the observer.  It is clear that in this case $\partial_\mu (\mathcal{H}_{sg}) = 0$, so that the Stern--Gerlach force vanishes. To calculate the Mathisson force in this case, we need to find the Riemann curvature tensor for the G\"odel-type universe since the Mathisson force is directly proportional to spacetime curvature.

In metric~\eqref{GT}, the nonzero components of the Riemann tensor can be obtained from 
\begin{equation}\label{V5} 
R_{0101} = \Omega^2,\  R_{0202} = (\kappa + \sigma)e^{2\mu x}\Omega^2,\ R_{0112} = -\,\sqrt{\sigma}e^{\mu x}
\Omega^2,\ R_{1212} = -\,\kappa\Bigl({\frac {4\kappa}\sigma} + 5\Bigr)e^{2\mu x}\Omega^2\,.
\end{equation}
We are interested in the components of the curvature tensor projected onto the orthonormal tetrad frame $\lambda^{\mu}{}_{\hat \alpha}$ adapted to our fiducial observer, namely, 
\begin{equation}\label{V6}
R_{\hat \alpha \hat \beta \hat \gamma \hat \delta} = R_{\mu \nu \rho \sigma}\,\lambda^{\mu}{}_{\hat \alpha}\,
\lambda^{\nu}{}_{\hat \beta}\,\lambda^{\rho}{}_{\hat \gamma}\,\lambda^{\sigma}{}_{\hat \delta}\,.
\end{equation}
The measured components of the Riemann tensor can be expressed via the symmetries of the Riemann tensor  as a $6\times 6$ matrix in the standard manner  with indices that range over the set $\{01,02,03,23,31,12\}$. The end result is of the general form  
\beq\label{V7}
\left[
\begin{array}{cc}
\mathbb {E} & \mathbb{H}\cr
\mathbb{H}^T & \mathbb{S}\cr 
\end{array}
\right]\,,
\eeq
where  $\mathbb{E}$, $\mathbb{H}$ and $\mathbb{S}$ represent the measured gravitoelectric,  gravitomagnetic and spatial components of the Riemann curvature tensor, respectively, and $\mathbb{E}$ and $\mathbb{S}$ are symmetric matrices, while $\mathbb{H}$ is traceless. In the case of G\"odel-type spacetime (\ref{GT}), we find $\mathbb{H}= 0$ and 
\begin{equation}\label{V8}
(\mathbb{E}_{\hat i \hat j}) = 
\begin{bmatrix}
\Omega^2&0&0 \\
0&\Omega^2&0 \\
0&0&0 
\end{bmatrix}
\,, \qquad (\mathbb{S}_{\hat i \hat j}) =
\begin{bmatrix}
0&0&0 \\
0&0&0 \\
0&0&-\left(1 + {\frac {4\kappa}{\sigma}}\right)\Omega^2  
\end{bmatrix}
\,.
\end{equation} 
These results are equally valid if the curvature tensor is projected onto the natural frame $e^{\mu}{}_{\hat \alpha}$ of the reference observer.

We find that the Mathisson force, given by Eq.~\eqref{M6}, can be expressed as
\begin{equation}\label{V9} 
\mathcal{F}^\mu = \lambda^{\mu}{}_{\hat \alpha}\,\mathcal{F}^{\hat \alpha}\,, \qquad \mathcal{F}^{\hat 0} = 0\,, \qquad \mathcal{F}^{\hat i}  = \mathbb{H}^{\hat i \hat j} S_{\hat j}\,.
\end{equation}
However, $\mathbb{H}= 0$; therefore,  the measured components of the Mathisson force vanish as well. That is, 
\begin{equation}\label{V10} 
\mathcal{F}_\mu = - \,\partial_\mu (\mathcal{H}_{sg}) = 0\,.
\end{equation}

It is important to verify this result in a quasi-inertial Fermi normal coordinate system established about the world line of an arbitrary reference observer that is spatially at rest. 


\section{Fermi Coordinates in G\"odel-type Spacetimes}\label{FermiG}

To explore spin-gravity coupling in Fermi coordinates, it is convenient to set up a quasi-inertial system of coordinates based on the nonrotating spatial frame adapted to a fiducial geodesic observer that is at rest in space with fixed $(x, y, z)$ coordinates and 4-velocity vector $u^\mu= \delta^\mu_0$ in G\"odel-type spacetime (\ref{GT}). The reference observer establishes in the neighborhood of its world line a  Fermi normal coordinate system based on the parallel-propagated spatial frame $\lambda^{\mu}{}_{\hat i}$, $i = 1,2,3$, given by Eq.~\eqref{G10}. That is, at each event $\bar{x}^\mu(\tau)$ on its world line, there is a local hypersurface formed by all spacelike geodesic curves that are orthogonal to the observer's world line at $\bar{x}^\mu(\tau)$. Consider an event with coordinates $x^\mu$ on this hypersurface that can be connected to $\bar{x}^\mu(\tau)$ by a unique spacelike geodesic of proper length $\ell$. Then, the reference observer can assign Fermi coordinates $X^{\mu} = (T,  X^i)$ to  $x^\mu$ such that
\begin{equation}\label{F1}
 T := \tau\,, \qquad X^i :=  \ell\, \xi^\mu \lambda_{\mu}{}^{\hat i}(\tau)\,.
\end{equation}
Here, $\xi^\mu$, $\xi^\mu\,u_\mu = 0$, is a unit spacelike vector tangent to the unique spacelike geodesic at $\bar{x}^\mu(\tau)$.

For the case of G\"odel's universe, one can find the exact Fermi metric coefficients~\cite{Chicone:2005vn}. The previous results are generalized here for G\"odel-type spacetime (\ref{GT}). For the spacelike geodesics $x^\mu(\ell)$, we use  Killing vector fields $\partial_t, \partial_y$ and $\partial_z$ to derive the equations of motion
\begin{align}\label{geo1}
t' + \sqrt{\sigma}\,e^{\mu x}\,y' = E\,,\quad \sqrt{\sigma}\,e^{\mu x}\,t' - \kappa\,e^{2\mu x}\,y' &= k\,,\\
z' = h\,,\quad -\,t'^2 -2\sqrt{\sigma}\,e^{\mu x}\,t'\,y' + x'^2 + \kappa e^{2\mu x}\,y'^2 + z'^2 &= 1\,.\label{geo2}
\end{align}
Here, $E, k$ and $h$ are integration constants; moreover,  a prime denotes the derivative of a spacetime coordinate with respect to proper length $\ell$, e.g., $t' = dt/d\ell$. The condition  $\xi_\mu \lambda^{\mu}{}_{\hat 0} = 0$, where $\xi^\mu = dx^\mu / d\ell$, implies $E = 0$. Then, with $z = h\,\ell$ and $E = 0$, we find
\begin{align}
t' &= {\frac {\sqrt{\sigma}\,e^{-\,\mu x}\,k}{\sigma + \kappa}}\,,\label{geo3}\\
y' &= -\,{\frac {e^{-\,2\mu x}\,k}{\sigma + \kappa}}\,,\label{geo4}\\
x'^2 &+ {\frac {e^{-\,2\mu x}\,k^2}{\sigma + \kappa}} = 1 - h^2\,.\label{geo5}
\end{align}
Ordinary differential equation (\ref{geo5}) has the general solution for $x(\ell)$ given by
\begin{equation}
e^{\mu x} = \alpha_0\,\cosh\left(a\ell + b\right),\label{xsol}
\end{equation}
where the constant parameters are fixed as
\begin{equation}\label{aa}
\alpha_0\,a = {\frac {|k|\,\mu}{\sqrt{\sigma + \kappa}}}\,,\qquad a = \mu\sqrt{1 - h^2}\,
\end{equation}
and the condition
\begin{equation}\label{ab}
\alpha_0\,\cosh b = 1
\end{equation}
is imposed to satisfy $x(0) = 0$.

Substituting Eq.~\eqref{xsol} into Eqs.~\eqref{geo3} and~\eqref{geo4}, we find the solutions for $t(\ell)$ and $y(\ell)$:
\begin{align}
t - \tau &= {\frac {2}{\mu}}\,\sqrt{\frac {\sigma}{\sigma + \kappa}}\,{\frac {k}{|k|}}\,\left[
\arctan e^{a\ell + b} - \arctan e^b\right]\,,\label{tsol}\\
y &= -\,{\frac {1}{\mu\,\alpha_0}}\,{\frac 1{\sqrt{\sigma + \kappa}}}\,{\frac {k}{|k|}}\,
\left[\tanh\,(a\ell + b) - \tanh b\right]\,.\label{ysol}
\end{align}
Then, making use of Eqs.~\eqref{G9}--\eqref{G10} and~\eqref{F1}, we derive for the Fermi coordinates
\begin{align}
T = \tau\,,\qquad Z &= \ell\,h\,,\label{TZ}\\ \label{XY1}
X\cos(\Omega T) - Y\sin(\Omega T) &= {\frac {\ell\,|k|}{\sqrt{\sigma + \kappa}}}\,\sinh b\,,\\
X\sin(\Omega T) + Y\cos(\Omega T) &= -\,{\frac {\ell\,k}{\sqrt{\sigma + \kappa}}}\,.\label{XY2}
\end{align}
As in~\cite{Chicone:2005vn}, we introduce cylindrical coordinates
\begin{equation}\label{XY}
X = \rho\cos\theta\,,\qquad Y = \rho\sin\theta\,
\end{equation}
and recast Eqs.~\eqref{XY1}--\eqref{XY2} into
\begin{align}
\cos(\theta + \Omega T) &= \tanh b\,,\label{cos}\\
\sin(\theta + \Omega T) &= -\,{\frac {|k|}{k\,\cosh b}}\,,\label{sin}\\
\mu\rho &= \ell\,a\,.\label{rho}
\end{align}
As a result, we rewrite the solutions (\ref{xsol}), (\ref{ysol}) and (\ref{tsol}) as 
\begin{align}
e^{\mu x} &= \cosh(\mu\rho) + \sinh(\mu\rho)\,\cos(\theta + \Omega T)\,,\label{x1}\\
\sqrt{\sigma + \kappa}\,\mu\,y &= {\frac {\tanh(\mu\rho)\,\sin(\theta + \Omega T)}
{1 + \tanh(\mu\rho)\,\cos(\theta + \Omega T)}}\,,\label{y1}\\
\tan\left[{\frac \mu 2}\sqrt{\frac {\sigma + \kappa} \sigma}\left(T - t\right)\right] &=
{\frac {\left(e^{\mu\rho} - 1\right)\sin(\theta + \Omega T)}
{1 - \cos(\theta + \Omega T) + [1 + \cos(\theta + \Omega T)]\,e^{\mu\rho}}}\,.\label{t1}
\end{align}

Finally, the transformation from $(t, x, y, z)$ to Fermi coordinates $(T, X, Y, Z)$ can be conveniently written in terms of the new variables
\begin{equation}
\mathfrak{R} = \mu\rho\,,\qquad \mathfrak{F} = \theta + \Omega T\,,\label{uv}
\end{equation}
as follows:
\begin{align}
e^{\mu x} &= \cosh \mathfrak{R} + \sinh\mathfrak{R}\,\cos\mathfrak{F}\,,\label{x2}\\
\sqrt{\sigma + \kappa}\,\mu\,y &= {\frac {\sinh\mathfrak{R}\,\sin\mathfrak{F}}
{\cosh\mathfrak{R} + \sinh\mathfrak{R}\,\cos\mathfrak{F}}}\,,\label{y2}\\
\tan\left[{\frac \mu 2}\sqrt{\frac {\sigma + \kappa} \sigma}\left(T - t\right)\right] &=
{\frac {\left(e^\mathfrak{R} - 1\right)\sin\mathfrak{F}}
{1 - \cos\mathfrak{F} + (1 + \cos\mathfrak{F})\,e^\mathfrak{R}}}\,.\label{t2}
\end{align}

By differentiation, we get
\begin{align}
dt + \sqrt{\sigma}\,e^{\mu x}dy &= dT + {\frac 1\mu}\sqrt{\frac \sigma {\sigma + \kappa}}
\,(\cosh\mathfrak{R} - 1)\,d\mathfrak{F}\,,\label{dt}\\
dx^2 + (\kappa + \sigma)\,e^{2\mu x}\,dy^2 &= {\frac 1{\mu^2}}\left(d\mathfrak{R}^2
+ \sinh^2\!\mathfrak{R}\,d\mathfrak{F}^2\right)\,.\label{dx}
\end{align}
It remains to substitute these results into Eq.~\eqref{GT1} to derive the line element of the G\"odel-type universe in terms of Fermi coordinates. We find
\begin{align}
ds^2 =& -\,(1 + \mathbb{L})\,dT^2 - 2\Omega\,\mathbb{K}\,dT\,(XdY - YdX) \nonumber\\
& + dX^2 + dY^2 + dZ^2 + {\frac {\mathbb{F}}{X^2 + Y^2}}(XdY - YdX)^2\,,\label{GTF}
\end{align}
where 
\begin{align}
\mathbb{L} &= {\frac {\sigma}{4(\sigma + \kappa)}}\left[\sinh^2\mathfrak{R}
- {\frac {\sigma + 2\kappa}{\sigma + \kappa}}(\cosh\mathfrak{R} - 1)^2\right]\,,\label{LF}\\
\mathbb{K} &= -\,{\frac {\kappa}{\sigma + \kappa}}\,{\frac {(\cosh\mathfrak{R} - 1)^2}{\mathfrak{R}^2}}\,,\label{FF}\\
\mathbb{F} &= {\frac {\sinh^2\mathfrak{R}}{\mathfrak{R}^2}} - 1 - {\frac {\sigma}{\sigma + \kappa}}\,
{\frac {(\cosh\mathfrak{R} - 1)^2}{\mathfrak{R}^2}}\,\label{HF}
\end{align}
are functions of the variable
\begin{equation}\label{F11}
\mathfrak{R} = 2\Omega\,\sqrt{\frac {\sigma + \kappa}{\sigma}}\,(X^2+Y^2)^{1/2}\,.
\end{equation}

\section{Spin-Gravity Coupling in Fermi Coordinates}\label{FermiC}

In general, the spacetime metric in the Fermi system is given by
\begin{equation}\label{F2}
 ds^2 = \hat{g}_{\mu \nu}\,dX^{\mu} dX^{\nu}\,,
\end{equation}
where
\begin{equation}\label{F3}
\hat{g}_{0 0} = - 1 - R_{\hat 0 \hat i \hat 0 \hat j}(T)\,X^i X^j+\cdots\,,\quad \hat{g}_{0 i} =  - \frac{2}{3}\,R_{\hat 0 \hat j \hat i \hat k}(T)\,X^j X^k + \cdots\,
\end{equation}
and
\begin{equation}\label{F4}
\hat{g}_{i j} = \delta_{ij} - \frac{1}{3}\,R_{\hat i \hat k \hat j \hat l}(T)\,X^k X^l + \cdots\,.
\end{equation}
In these expansions in powers of spatial Fermi coordinates, the coefficients are in general functions of $T$ and consist of components of Riemann curvature tensor and its covariant derivatives as measured by the reference observer that permanently occupies the spatial origin of Fermi coordinate system.  That is, the metric of the Fermi normal coordinate system established on the basis of a parallel-propagated spatial frame along the world line of a geodesic observer is the Minkowski metric plus perturbations caused by the curvature of spacetime. 
Fermi coordinates are admissible within a cylindrical spacetime region around the world line of the fiducial observer and the radius of this cylinder is given by an appropriate radius of curvature of spacetime~\cite{Chicone:2005vn}. 

As defined in Eq.~\eqref{V6},  $R_{\hat \alpha \hat \beta \hat \gamma \hat \delta}$ are evaluated at the origin of spatial Fermi coordinates via the projection of the Riemann tensor on the tetrad frame $\lambda^{\mu}{}_{\hat \alpha}$ of the fiducial observer; indeed, for the stationary G\"odel-type spacetime the nonzero components of $R_{\hat \alpha \hat \beta \hat \gamma \hat \delta}$ are constants and can be obtained from 
\begin{equation}\label{F5}
R_{\hat 0 \hat 1 \hat 0 \hat 1} = R_{\hat 0 \hat 2 \hat 0 \hat 2} = \Omega^2,\qquad
R_{\hat 1 \hat 2 \hat 1 \hat 2} = -\,\left(1 + {\frac {4\kappa}{\sigma}}\right)\Omega^2\,
\end{equation}
via the symmetries of the Riemann curvature tensor. 

We define the curvature-based gravitoelectric  potential $\hat{\Phi}$ and gravitomagnetic vector potential $\hat{\bm{A}}$ via $\hat{g}_{0 0} = -1 + 2 \hat{\Phi}$ and $\hat{g}_{0 i} = - 2 \hat{A}_i$~\cite{Mashhoon:2003ax, Bini:2021gdb}. Indeed,  
\begin{equation}\label{F6}
\hat{\Phi} = -\frac{1}{2}\, R_{\hat 0 \hat i \hat 0 \hat j}\,X^i X^j + \cdots\,, \qquad \hat{A}_i =  \frac{1}{3}\,R_{\hat 0 \hat j \hat i \hat k}\,X^j X^k+\cdots\,.
\end{equation}
The corresponding fields are given by
\begin{equation}\label{F7}
\hat{\bm{E}} = - \bm{\nabla} \hat{\Phi}\,, \qquad \hat{\bm{B}} =  \bm{\nabla} \times \hat{\bm{A}}\,,
\end{equation}
 as expected; more explicitly, 
\begin{equation}\label{F8}
\hat{E}_i = R_{\hat 0 \hat i \hat 0 \hat j}\,X^j+\cdots\,, \qquad \hat{B}_i =  -\frac{1}{2}\,\epsilon_{ijk}\,R^{\hat j \hat k}{}_{\hat 0 \hat l}\,X^l+\cdots\,.
\end{equation}
To lowest order, the gravitomagnetic field vanishes in the G\"odel-type spacetime; therefore, we need to compute higher-order terms. 

In Section~\ref{FermiG}, we derived the exact Fermi metric coefficients for the G\"odel-type universe. They are given explicitly by
\begin{align}\label{F9}
\hat{g}_{0 0} &= 
-\,1 - {\frac {\sigma}{4(\sigma + \kappa)}}\left[\sinh^2\mathfrak{R}
- {\frac {\sigma + 2\kappa}{\sigma + \kappa}}(\cosh\,\mathfrak{R} - 1)^2\right],\\
\hat{g}_{0 i} &= -\,{\frac {\kappa}{\sigma + \kappa}}\,\Omega\,\frac{(\cosh \mathfrak{R} - 1)^2}
{\mathfrak{R}^2} (Y, -X, 0)\,\label{F9a}
\end{align}
and
\begin{equation}\label{F10}
(\hat{g}_{i j}) =
\begin{bmatrix}
1+ \mathbb{A}&- \,\mathbb{C}&0 \\
- \,\mathbb{C}&1+ \mathbb{B}&0 \\
0&0&1 
\end{bmatrix}\,,
\end{equation}
where
\begin{equation}\label{F12}
\mathbb{A} = \mathbb{F} \,\frac{Y^2}{X^2+Y^2}\,, \qquad  \mathbb{B} = \mathbb{F} \,\frac{X^2}{X^2+Y^2}\,, \qquad \mathbb{C} = \mathbb{F}\, \frac{XY}{X^2+Y^2}\,.
\end{equation}
The exact Fermi coordinate system has been established around the fiducial observer fixed at $X=Y=Z=0$.

For $\kappa\ge 0$, there are no closed timelike curves. In the special case of the G\"odel universe with parameters (\ref{msk}), there are no closed timelike curves within a cylindrical region about the $Z$ axis with
\begin{equation}\label{F14}
\mathfrak{R} = \sqrt{2}\,\Omega\, (X^2+Y^2)^{1/2} \leq \mathfrak{R}_{\rm max}\,,\qquad   \mathfrak{R}_{\rm max} = 2 \ln (1+\sqrt{2})\,.
\end{equation}
Indeed,  a circle in the $(X, Y)$ plane inside this domain is spacelike; however, it becomes null for $\mathfrak{R} = \mathfrak{R}_{\rm max}$ and timelike for $\mathfrak{R} > \mathfrak{R}_{\rm max}$.

The stationary and divergence-free gravitomagnetic vector field of the G\"odel-type universe is given by $\hat{B}_1 = \hat{B}_2 = 0$ and 
\begin{equation}\label{F15}
\hat{B}_3 = -\,{\frac {\kappa}{\sigma + \kappa}}\,\Omega (\cosh \mathfrak{R} -1) \frac{\sinh \mathfrak{R}}{\mathfrak{R}}\,.
\end{equation}
It is interesting to note that $\hat{B}_3$ and its first derivative with respect to $\mathfrak{R}$ vanish at $\mathfrak{R} = 0$; then, $\hat{B}_3$ monotonically increases with increasing $\mathfrak{R}$ and diverges as $\mathfrak{R} \to \infty$. More explicitly,
\begin{equation}\label{F16}
\hat{B}_3 = -\,{\frac {2\kappa}{\sigma}}\Omega^3 (X^2+Y^2)\left[1 + \Omega^2
\Bigl({\frac {\sigma + \kappa}{\sigma}}\Bigr)(X^2+Y^2) + {\frac{2\Omega^4}{5}}
\Bigl({\frac {\sigma + \kappa}{\sigma}}\Bigr)^2(X^2+Y^2)^2 + \cdots\right]\,,
\end{equation}
so that the fiducial observer measures a null gravitomagnetic field at its location $(X=Y=Z=0)$. Furthermore, the gravitomagnetic field away from the $Z$ axis points along $Z$ and is cylindrically symmetric; indeed, it vanishes all along $Z$, but increases monotonically away from the $Z$ axis and eventually diverges as the radius of the cylinder about the $Z$ axis approaches infinity.  

Within the Fermi coordinate system, it is useful to define the class of fundamental observers that remain at rest in space, each with fixed $(X, Y, Z)$ coordinates. For our present purposes, we concentrate on the set of fundamental  observers that occupy a cylindrical region in the neighborhood of the $Z$ axis. Specifically, in this region we can express the metric tensor in Fermi coordinates  as  
\begin{equation}\label{F17}
\hat{g}_{\mu \nu} = \eta_{\mu \nu} + \hat{h}_{\mu \nu}\,,
\end{equation}
where the nonzero components of the gravitational potentials are given by
\begin{equation}\label{F18}
\hat{h}_{0 0} = -\,\Omega^2 (X^2+Y^2)\,, \quad \hat{h}_{0 1} = -\,{\frac {\kappa}{\sigma}}\Omega^3 (X^2+Y^2) Y
\,, \quad \hat{h}_{0 2} = {\frac {\kappa}{\sigma}}\Omega^3 (X^2+Y^2) X\,,
\end{equation}
and
\begin{equation}\label{F19}
\hat{h}_{1 1} = \frac{1}{3}\Bigl(1 + {\frac {4\kappa}{\sigma}}\Bigr)\Omega^2Y^2\,,\quad
\hat{h}_{1 2} = -\,\frac{1}{3}\Bigl(1 + {\frac {4\kappa}{\sigma}}\Bigr)\Omega^2X Y\,,\quad
\hat{h}_{2 2} = \frac{1}{3}\Bigl(1 + {\frac {4\kappa}{\sigma}}\Bigr)\Omega^2 X^2\,.
\end{equation}
That is, for the sake of simplicity we confine our considerations  to a cylindrical region about the $Z$ axis such that $\Omega\, |X| = \Omega\, |Y| \lesssim \varepsilon$, where $0 < \varepsilon \ll 1$ and all terms of order $\varepsilon^4$ and higher are neglected in our analysis.

In the cylindrical neighborhood of the fiducial observer under consideration, fundamental observers have access to adapted orthonormal tetrad frames $\varphi^{\mu}{}_{\hat \alpha}$ given in the Fermi coordinate $(T, X, Y, Z)$ system by 
\begin{align}\label{F20}
\varphi^{\mu}{}_{\hat 0} &= ( 1 + \tfrac{1}{2}\hat{h}_{ 0 0}, 0, 0, 0)\,, \qquad \varphi^{\mu}{}_{\hat 1} = ( \hat{h}_{0 1}, 1 - \tfrac{1}{2}\hat{h}_{1 1}, 0, 0)\,,\\
\varphi^{\mu}{}_{\hat 2} &= ( \hat{h}_{0 2}, -\hat{h}_{1 2}, 1 - \tfrac{1}{2}\hat{h}_{2 2}, 0)\,, \qquad \varphi^{\mu}{}_{\hat 3} = ( 0, 0, 0, 1)\,.\label{F21}
\end{align}
These tetrad axes are primarily along the Fermi coordinate directions; indeed, for $X = Y = 0$, $\varphi^{\mu}{}_{\hat \alpha} \to \lambda^{\mu}{}_{\hat \alpha}$. According to these fundamental observers, a spinning particle held at rest in space has a 4-velocity vector in the Fermi system given by $\hat{U}^\mu = \varphi^{\mu}{}_{\hat 0}$; moreover, its spin vector has measured components
\begin{equation}\label{F22}
\hat{S}_{\hat 0} = 0\,, \qquad \hat{S}_{\hat i} = \hat{S}_\mu\, \varphi^{\mu}{}_{\hat i}\,,
\end{equation}
since $\hat{S}^\mu\,\hat{U}_\mu = 0$. Furthermore, the gravitomagnetic field at the location of the spin is given by 
\begin{equation}\label{F23}
\hat{B}_1 = 0\,, \qquad \hat{B}_2 = 0\,, \qquad \hat{B}_3 = -\,\frac{1}{2} (\partial_X\,\hat{h}_{0 2}
- \partial_Y\,\hat{h}_{0 1}) = -\,{\frac {2\kappa}{\sigma}}\Omega^3 (X^2+Y^2)\,,
\end{equation}
in agreement with Eq.~\eqref{F16} within our approximation scheme. The Hamiltonian for spin-gravity coupling in the Fermi frame is thus given by 
\begin{equation}\label{F24}
\mathcal{\hat{H}}_{sg} = \bm{\hat{S}}\cdot\bm{\hat{B}} =
-\,{\frac {2\kappa}{\sigma}}\Omega^3 (X^2+Y^2)\hat{S}^{\hat 3}\,,
\end{equation}
which reduces in our approximation to $-\,{\frac {2\kappa}{\sigma}}\Omega^3 (X^2+Y^2)S^{\hat 3}$, where $S^{\hat 3}$ is a constant. The corresponding Stern--Gerlach force is then
\begin{equation}\label{F25}
-\,\partial_\mu\,\mathcal{\hat{H}}_{sg}  = {\frac {4\kappa}{\sigma}}\,\Omega^3\,S^{\hat 3}(0, X, Y, 0)\,.
\end{equation}

Next, we need to compute the Mathisson force in the Fermi frame, namely, 
\begin{equation}\label{F26}
\hat{\mathcal{F}}_\mu  = - \,\frac{1}{2}\,\hat{R}_{\mu \nu \alpha \beta} \hat{U}^{\nu}\,\hat{S}^{\alpha \beta}\,.
\end{equation}
For metric~\eqref{F17}, the curvature tensor to first order in the perturbation is given by
\begin{equation}\label{F27}
\hat{R}_{\mu \nu \alpha \beta} =  \frac{1}{2}\,(\hat{h}_{\mu \beta, \, \nu \alpha}  + \hat{h}_{\nu \alpha, \,\mu \beta} - \hat{h}_{\nu \beta, \, \mu \alpha} - \hat{h}_{\mu \alpha, \, \nu \beta})\,.
\end{equation}
We are interested in the gravitomagnetic components of this curvature tensor as measured by the fundamental observers. Projection of this tensor on the tetrad frame $\varphi^{\mu}{}_{\hat \alpha}$ does not affect its components in our approximation scheme. We find in this case
\begin{equation}\label{F28}
(\mathbb{\hat{H}}_{\hat i \hat j}) = 
\begin{bmatrix}
0&0&\kappa_1 \\
0&0&\kappa_2 \\
0&0&0 
\end{bmatrix}
\,,
\end{equation} 
where
\begin{equation}\label{F29}
\kappa_1 = \frac{1}{2} \partial_X\,(\partial_X\,\hat{h}_{0 2}  - \partial_Y\,\hat{h}_{0 1})\,,\qquad \kappa_2 = \frac{1}{2} \partial_Y\,(\partial_X\,\hat{h}_{0 2}  - \partial_Y\,\hat{h}_{0 1})\,.
\end{equation}
Hence, $\hat{\mathcal{F}}_{\hat 0} = 0$ and $\hat{\mathcal{F}}_{\hat i}  = \mathbb{\hat{H}}_{\hat i \hat j} \hat{S}^{\hat j} = (\kappa_1, \kappa_2, 0) S^{\hat 3}$ at the level of approximation under consideration here. Moreover, Eq.~\eqref{F23} implies
\begin{equation}\label{F30}
\kappa_1  = {\frac {4\kappa}{\sigma}}\,\Omega^3 X\,,\qquad \kappa_2 =
{\frac {4\kappa}{\sigma}}\,\Omega^3 Y\,.
\end{equation}
Therefore, $\hat{\mathcal{F}}_\mu  = -\,\partial_\mu\,\mathcal{\hat{H}}_{sg}$ as measured by the fundamental observers within the cylindrical domain in the Fermi frame. 

We have thus far relied on the classical pole-dipole model for the evaluation of spin-gravity coupling. It is important to demonstrate that our considerations are consistent with the solutions of the Dirac equation in the G\"odel-type universe.


\section{Dirac Equation in the G\"odel-type Universe}\label{Dirac}

Let us start with the Dirac equation in the form~\cite{Obukhov:2013zca, Obukhov:2017avp}
\begin{equation}\label{H1}
(i\gamma^\alpha \nabla_\alpha  - m) \, \Psi = 0\,, \qquad \nabla_\mu = \partial_\mu + \Gamma_\mu\,,
\end{equation}
where the fermion wave function $\Psi$ is a 4-component spacetime scalar variable composed of the pair of 2-spinors $\varphi$ and $\chi$:
\begin{equation}\label{s1}
\Psi = \begin{bmatrix} \varphi \\ \chi \end{bmatrix},\qquad 
\varphi = \begin{bmatrix} \varphi_1 \\ \varphi_2\end{bmatrix},\qquad
\chi = \begin{bmatrix}\chi_1 \\ \chi_2 \end{bmatrix}\,.
\end{equation}
As before, we assume the observer in the gravitational field has a natural adapted orthonormal tetrad field and 
\begin{equation}\label{H2}
\gamma^\alpha = e^{\alpha}{}_{\hat \beta}\,\gamma^{\hat \beta}\,,\qquad \{\gamma^\mu, \gamma^\nu\} = - 2 g^{\mu \nu}(x) I_4\,,
\end{equation}
where $I_n$ is the $n$-dimensional identity matrix and 
\begin{equation}\label{H3}
\gamma^{\hat 0} = 
\begin{bmatrix}
I_2&0 \\
0&-I_2 
\end{bmatrix}
\,, \qquad 
\gamma^{\hat i} = 
\begin{bmatrix}
0&\sigma_i \\
-\sigma_i&0 
\end{bmatrix}
\,.
\end{equation}
Here, $\sigma_i$ are Pauli matrices, namely, 
\begin{equation}\label{H4}
\sigma_1 = 
\begin{bmatrix}
0&1 \\
1&0 
\end{bmatrix}
\,, \qquad 
\sigma_2 = 
\begin{bmatrix}
0&-i \\
i&0 
\end{bmatrix}
\,, \qquad
\sigma_3 = 
\begin{bmatrix}
1&0 \\
0&-1 
\end{bmatrix}
\,.
\end{equation}

The spin connection $\Gamma_\mu$ (also known as Fock-Ivanenko coefficients) is given by
\begin{equation}\label{H7}
\Gamma_\mu = -\,\frac{i}{4} \, e^{\nu}{}_{\hat \alpha}\, e_{\nu \,\hat \beta ; \mu}\, \sigma^{\hat \alpha \hat \beta}\,, \qquad \sigma^{\hat\alpha \hat \beta} := \frac{i}{2} [\gamma^{\hat \alpha}, \gamma^{\hat \beta}]\,.
\end{equation}
Making use of tetrad frame (\ref{G9}), we find, after some algebra, the explicit form of the Dirac equation (\ref{H1}) in  G\"odel-type spacetime (\ref{GT}):
\begin{align}
\Bigl[\Bigl(\gamma^{\hat 0} - \sqrt{\frac {\sigma}{\sigma + \kappa}}\,\gamma^{\hat 2}\Bigr)i\partial_t
- \gamma^{\hat 1}\,p_x - {\frac {e^{-\mu x}}{\sqrt{\kappa + \sigma}}}\,\gamma^{\hat 2}\,p_y
- \gamma^{\hat 3}\,p_z\nonumber\\
+\, {\frac {i\mu}{2}}\gamma^{\hat 1} + {\frac \mu 4}\sqrt{\frac {\sigma}{\sigma + \kappa}}
\,\gamma^{\hat 0}\,\Sigma^{\hat 3} - m\Bigr]\Psi = 0\,.\label{DirL}
\end{align}
Here, as usual, the momentum operator is $\bm{p} = -\,i\bm{\nabla}$ and the spin operator $\bm{\Sigma}$ is given by the matrix
\begin{equation}\label{spinS}
\Sigma^{\hat i} = 
\begin{bmatrix}
\sigma_i & 0 \\
0 & \sigma_i 
\end{bmatrix}\,.
\end{equation}

Next, due to the symmetries of G\"odel-type spacetime, we assume a solution of the form
\begin{equation}\label{H10}
\Psi = \psi(x) \exp (-i\,\omega\, t + i\,k_2\, y +i\, k_3\,z)\,,
\end{equation}
where the four components of $\psi(x)$ satisfy ordinary differential equations, namely,  
\begin{equation}\label{H11}
\frac{d\psi}{dx} = \mathcal{M} \psi\,, 
\end{equation}
where  $\mathcal{M}$ is the $4\times 4$ matrix
\begin{equation}\label{H12}
\mathcal{M} =
\begin{bmatrix}
\mathcal{A}_{+}&ik_3&0&i\mathcal{B}_{+} \\
-ik_3&-\mathcal{A}_{-}&i\mathcal{B}_{-}&0  \\
0&i\mathcal{B}_{+}&\mathcal{A}_{+}&ik_3 \\
i\mathcal{B}_{-}&0&-ik_3&-\mathcal{A}_{-}
\end{bmatrix} + im\begin{bmatrix} 0&0&0&1 \\
0&0&1&0  \\ 0&-1&0&0 \\ -1&0&0&0 \end{bmatrix}\,.
\end{equation}
Here, $\mathcal{A}_{\pm}$ and $\mathcal{B}_{\pm}$ are given by
\begin{equation}\label{H13}
\mathcal{A}_{\pm} =  \omega\sqrt{\frac {\sigma}{\sigma + \kappa}} \pm \Omega
\,\sqrt{\frac {\sigma + \kappa}{\sigma}} + k_2 \,e^{-\,\mu x}\,,
\qquad \mathcal{B}_{\pm} =  \omega \pm {\frac{\Omega}2}\,.
\end{equation}
The spin-vorticity-gravity coupling is evident in the way the frequency of the radiation is changed by $\pm \Omega/2$ in agreement with previous results~\cite{Bini:2021gdb, Mashhoon:2013jaa, Bini:2012ht}. If $k_2 = 0$, the waves can only travel parallel or antiparallel to the rotation axis. In this case, matrix $\mathcal{M}$ has constant elements and the general solution of Eq.~\eqref{H11} can be expressed in terms of the eigenvalues and eigenfunctions of $\mathcal{M}$.  It turns out that no propagation can occur in this case due to the requirement that the wave amplitude be finite at all times~\cite{Bini:2012ht}. These general results for the Dirac equation are consistent with the propagation of scalar and electromagnetic waves in the G\"odel-type universe; for brief accounts of these latter topics, see the appendices at the end of this paper.  

To deal with the general case, we henceforth assume $k_2 \ne 0$ and change to $\xi = e^{-\,\mu x}$ instead of $x$ as the independent variable. Let us recall here that $\mu >0$, since we have explicitly assumed $\Omega > 0$.  For $\infty > x > -\infty$, we find $\xi$ goes from zero to $+\infty$; hence, $\xi$ is a radial coordinate. In terms of $\xi$, Eq.~\eqref{H11} takes the form
\begin{equation}\label{H14}
\xi\, \frac{d\psi}{d\xi} = \mathbb{M} \psi\,, 
\end{equation}
where  matrix $\mathbb{M}$ is simply related to $\mathcal{M}$, namely,
\begin{equation}\label{H15}
\mathbb{M} =
\begin{bmatrix}
-\bar{\mathcal{A}}_{+}&-i\gamma&0&-i\bar{\mathcal{B}}_{+} \\
i\gamma&\bar{\mathcal{A}}_{-}&-i\bar{\mathcal{B}}_{-}&0  \\
0&-i\bar{\mathcal{B}}_{+}&-\bar{\mathcal{A}}_{+}& -i\gamma \\
-i\bar{\mathcal{B}}_{-}&0&i\gamma&\bar{\mathcal{A}}_{-}
\end{bmatrix} - {\frac {im}{2\Omega}}\,\sqrt{\frac {\sigma}{\sigma + \kappa}}
\begin{bmatrix} 0&0&0&1 \\ 0&0&1&0  \\ 0&-1&0&0 \\ -1&0&0&0 \end{bmatrix}
\,. 
\end{equation}
Here, $\bar{\mathcal{A}}_{\pm}$ and $\bar{\mathcal{B}}_{\pm}$ are given by
\begin{equation}\label{H16}
\bar{\mathcal{A}}_{\pm} =  \frac{\omega}{2\Omega}\left({\frac {\sigma}{\sigma + \kappa}}\right)
\pm \frac{1}{2} + \beta\,\xi\,, \qquad \bar{\mathcal{B}}_{\pm} = \frac{1}{2}
\,\sqrt{\frac {\sigma}{\sigma + \kappa}}\left(\frac{\omega}{\Omega} \pm \frac{1}{2}\right)\,
\end{equation}
and we have introduced dimensionless parameters  
\begin{equation}\label{H17}
\beta = \frac{k_2}{2\Omega}\,\sqrt{\frac {\sigma}{\sigma + \kappa}}\,,
\qquad \gamma = \frac{k_3}{2\Omega}\,\sqrt{\frac {\sigma}{\sigma + \kappa}}\,.
\end{equation}

To clarify the structure of the resulting system (\ref{H14})--(\ref{H15}), we note that the 4-spinor (\ref{s1}) can be decomposed into the sum of the left and right spinors,
\begin{equation}\label{s2}
\psi = \psi^L + \psi^R\,,\quad \psi^L = {\frac 12}(1 - \gamma_5)\psi\,,\quad \psi^R = {\frac 12}(1 + \gamma_5)\psi\,,
\end{equation}
where $\gamma_5 := i\,\gamma^{\hat 0}\gamma^{\hat 1}\gamma^{\hat 2}\gamma^{\hat 3}$. By definition, the left and right spinors are eigenstates of the $\gamma_5$ matrix: $\gamma_5\psi^L = -\,\psi^L$ and $\gamma_5\psi^R = \psi^R$. Furthermore, we decompose the left and right spinors into the eigenstates of the $\Sigma^{\hat 3}$ spin matrix (i.e., ``spin-up'' and ``spin-down'' states):
\begin{equation}\label{s3}
\psi^L = \psi^L_+ + \psi^L_-\,,\quad \psi^R = \psi^R_+ + \psi^R_-\,,\quad \Sigma^{\hat 3}\psi^L_{\pm} = \pm\psi^L_{\pm}\,,
\quad \Sigma^{\hat 3}\psi^R_{\pm} = \pm\psi^R_{\pm}\,. 
\end{equation}
After these steps, we thus have
\begin{equation}\label{s4}
\psi^L_+ = {\cal L}_+\!\begin{bmatrix} 1 \\ 0  \\ 1 \\ 0 \end{bmatrix},\quad
\psi^L_- = {\cal L}_-\!\begin{bmatrix} 0 \\ 1  \\ 0 \\ 1 \end{bmatrix},\quad
\psi^R_+ = {\cal R}_+\!\begin{bmatrix} 1 \\ 0  \\ -1 \\ 0 \end{bmatrix},\quad
\psi^R_- = {\cal R}_-\!\begin{bmatrix} 0 \\ 1  \\ 0 \\ -1 \end{bmatrix},
\end{equation}
where explicitly
\begin{align}\label{s5}
{\cal L}_+ = {\frac 12}(\varphi_1 + \chi_1)\,,\qquad {\cal L}_- = {\frac 12}(\varphi_2 + \chi_2)\,,\\ 
{\cal R}_+ = {\frac 12}(\varphi_1 - \chi_1)\,,\qquad {\cal R}_- = {\frac 12}(\varphi_2 - \chi_2)\,.\label{s6}
\end{align}

Taking these definitions into account, we can straightforwardly recast  system (\ref{H14})--(\ref{H15}) into an equivalent but more transparent form:
\begin{align}
\Bigl(\xi{\frac d{d\xi}} + \bar{\mathcal{A}}_{+}\Bigr){\cal L}_+ &= -\,i(\bar{\mathcal{B}}_{+} + \gamma)
{\cal L}_- +i{\frac {m}{2\Omega}}\,\sqrt{\frac {\sigma}{\sigma + \kappa}}\,{\cal R}_-\,,\label{l1}\\
\Bigl(\xi{\frac d{d\xi}} - \bar{\mathcal{A}}_{-}\Bigr){\cal L}_- &= -\,i(\bar{\mathcal{B}}_{-} - \gamma)
{\cal L}_+ + i {\frac {m}{2\Omega}}\,\sqrt{\frac {\sigma}{\sigma + \kappa}}\,{\cal R}_+\,,\label{l2}\\
\Bigl(\xi{\frac d{d\xi}} + \bar{\mathcal{A}}_{+}\Bigr){\cal R}_+ &=\quad i(\bar{\mathcal{B}}_{+} - \gamma)
{\cal R}_- -i {\frac {m}{2\Omega}}\,\sqrt{\frac {\sigma}{\sigma + \kappa}}\,{\cal L}_-\,,\label{r1}\\
\Bigl(\xi{\frac d{d\xi}} - \bar{\mathcal{A}}_{-}\Bigr){\cal R}_- &=\quad i(\bar{\mathcal{B}}_{-} + \gamma)
{\cal R}_+ - i {\frac {m}{2\Omega}}\,\sqrt{\frac {\sigma}{\sigma + \kappa}}\,{\cal L}_+\,.\label{r2}
\end{align}
The nontrivial mass mixes the left and right modes. However, for the massless ($m = 0$) case or in the high-energy approximation (${\frac {mc^2}{\hbar\Omega}} \ll 1$) we can neglect the last terms on the right-hand sides. As a result, the left modes ${\cal L}_\pm$ decouple from the right modes ${\cal R}_\pm$ and the system reduces to
\begin{align}
\Bigl(\xi{\frac d{d\xi}} + \bar{\mathcal{A}}_{+}\Bigr){\cal L}_+ &= -\,i(\bar{\mathcal{B}}_{+} + \gamma)
{\cal L}_-\,,\label{l01}\\
\Bigl(\xi{\frac d{d\xi}} - \bar{\mathcal{A}}_{-}\Bigr){\cal L}_- &= -\,i(\bar{\mathcal{B}}_{-} - \gamma)
{\cal L}_+\,,\label{l02}\\
\Bigl(\xi{\frac d{d\xi}} + \bar{\mathcal{A}}_{+}\Bigr){\cal R}_+ &=\quad i(\bar{\mathcal{B}}_{+} - \gamma)
{\cal R}_-\,,\label{r01}\\
\Bigl(\xi{\frac d{d\xi}} - \bar{\mathcal{A}}_{-}\Bigr){\cal R}_- &=\quad i(\bar{\mathcal{B}}_{-} + \gamma)
{\cal R}_+\,.\label{r02}
\end{align}

It is interesting to mention that in this approximation scheme Eq.~\eqref{H14} can also be solved by a different approach that is briefly described in Appendix~\ref{appA}. 

\subsection{Explicit solutions}

Multiplying Eq.~\eqref{l01} by $-\,i(\bar{\mathcal{B}}_{-} - \gamma)$ and Eq.~\eqref{l02} by $-\,i(\bar{\mathcal{B}}_{+} + \gamma)$, we derive the second-order equations for the {\it left modes}:
\begin{align}
\Bigl(\xi{\frac d{d\xi}} + \bar{\mathcal{A}}_{+}\Bigr)\Bigl(\xi{\frac d{d\xi}} - \bar{\mathcal{A}}_{-}\Bigr)
{\cal L}_- &= \left[-\,\bar{\mathcal{B}}_{+}\bar{\mathcal{B}}_{-} + \gamma(\bar{\mathcal{B}}_{+} -
\bar{\mathcal{B}}_{-}) + \gamma^2\right]{\cal L}_-\,,\label{L1}\\
\Bigl(\xi{\frac d{d\xi}} - \bar{\mathcal{A}}_{+}\Bigr)\Bigl(\xi{\frac d{d\xi}} + \bar{\mathcal{A}}_{-}\Bigr)
{\cal L}_+ &= \left[-\,\bar{\mathcal{B}}_{+}\bar{\mathcal{B}}_{-} + \gamma(\bar{\mathcal{B}}_{+} -
\bar{\mathcal{B}}_{-}) + \gamma^2\right]{\cal L}_+\,.\label{L2}
\end{align}
In Eq.~\eqref{H16}, it is useful to introduce a dimensionless parameter $\alpha$,
\begin{equation}\label{AA1}
\alpha := \frac{\omega}{2\Omega}\left({\frac {\sigma}{\sigma + \kappa}}\right)\,,\qquad \bar{\mathcal{A}}_{\pm} = \alpha  + \beta\,\xi \pm \frac{1}{2}\,;
\end{equation}
then, 
\begin{align}
\bar{\mathcal{A}}_{+}\bar{\mathcal{A}}_{-} = (\alpha + \beta\xi)^2 - {\frac 14}\,, &\qquad
\bar{\mathcal{A}}_{+} - \bar{\mathcal{A}}_{-} = 1\,,\label{AA2}\\
\bar{\mathcal{B}}_{+}\bar{\mathcal{B}}_{-} = {\frac {\sigma}{\sigma + \kappa}}\Bigl[
{\frac {\omega^2}{(2\Omega)^2}} - {\frac {1}{16}}\Bigr]\,, &\qquad
\bar{\mathcal{B}}_{+} - \bar{\mathcal{B}}_{-} = \frac{1}{2}
\,\sqrt{\frac {\sigma}{\sigma + \kappa}}\,.\label{BB2}
\end{align}
Employing the ansatz 
\begin{equation}\label{Lu}
{\cal L}_\pm = \xi^{-1}\,u_{\mp{\frac 12}}\,, 
\end{equation}
we can recast Eqs. (\ref{L1}) and (\ref{L2}) into the form
\begin{equation}\label{Du}
\xi^2{\frac {d^2}{d\xi^2}}u_s + \Bigl[{\frac 14} - \tilde{\mu}_f^2
- \beta^2\xi^2 - 2\beta\xi\,(\alpha + s)\Bigr]\,u_s = 0\,,
\end{equation}
where $s = \pm {\frac 12}$ and 
\begin{align}
\tilde{\mu}_f^2 &= \alpha^2 + \gamma^2 -\,\bar{\mathcal{B}}_{+}\bar{\mathcal{B}}_{-}
+ \gamma(\bar{\mathcal{B}}_{+} - \bar{\mathcal{B}}_{-})\nonumber\\
&= \frac{1}{\mu^2}\Bigl[ -\,\omega^2\,{\frac {\kappa}{\sigma + \kappa}}
+ \left(k_3 - \Omega/2\right)^2 \Bigr]\,.\label{muf}
\end{align}
With a new independent variable $\tilde{\xi} = 2 |\beta| \xi$, Eq. (\ref{Du}) can be reduced to Whittaker's equation~\cite{A+S}
\begin{equation}\label{H28}
\frac{d^2 u_s}{d\tilde{\xi}^2}  + \left[ -\,\frac{1}{4} + \frac{\tilde{\kappa}_f}{\tilde{\xi}}
+ \frac{\tfrac{1}{4} - \tilde{\mu}_f^2}{\tilde{\xi}^2}\right] u_s = 0\,,
\end{equation}
where
\begin{equation}\label{H29}
\tilde{\kappa}_f = -\,\frac{\beta}{|\beta|}\left(\alpha + s\right)\,. 
\end{equation}

The Dirac field is a linear perturbation on the G\"odel-type spacetime; therefore, $\psi(x)$ should be bounded. 
Demanding that $\psi(x)$ be finite everywhere, the acceptable solution of Whittaker's equation is given via the confluent hypergeometric functions by 
\begin{equation}\label{H30}
 u_s = u_s^0\, \exp(-\tfrac{1}{2} \tilde{\xi})\,\tilde{\xi}^{\tfrac{1}{2} + \tilde{\mu}_f}
\,{}_1F_1(\tfrac{1}{2} + \tilde{\mu}_f -\tilde{\kappa}_f, 1+2\tilde{\mu}_f; \tilde{\xi})\,,   
\end{equation}
where  
\begin{equation}\label{H31}
\frac{1}{2} + \tilde{\mu}_f -\tilde{\kappa}_f = -\,n\,, \quad n = 0, 1, 2, \dots\,.
\end{equation}
In this case, the confluent hypergeometric function can be expressed in terms of the associated Laguerre polynomial.  

For $k_2 < 0$, $\beta$ is negative and $\tilde{\kappa}_f = \alpha + s = \frac{\omega}{2\Omega}\left({\frac {\sigma}{\sigma + \kappa}}\right) + s$, with $s = \pm {\frac 12}$. Then, combining Eqs.~\eqref{H31} and (\ref{muf}), we derive the dispersion relation
\begin{equation}\label{disp}
\omega = (2n + 1 + 2s)\Omega \pm \left[ (k_3 - \Omega/2)^2
- \frac{\kappa}{\sigma}(2n + 1 + 2s)^2 \Omega^2 \right]^{1/2}\,.
\end{equation}
Note that solutions with both signs of energy are admissible. 

Similarly, multiplying Eq.~\eqref{r01} by $i(\bar{\mathcal{B}}_{-} + \gamma)$ and Eq.~\eqref{r02} by $i(\bar{\mathcal{B}}_{+} - \gamma)$, we derive the second-order equations for the {\it right modes}:
\begin{align}
\Bigl(\xi{\frac d{d\xi}} + \bar{\mathcal{A}}_{+}\Bigr)\Bigl(\xi{\frac d{d\xi}} - \bar{\mathcal{A}}_{-}\Bigr)
{\cal R}_- &= \left[-\,\bar{\mathcal{B}}_{+}\bar{\mathcal{B}}_{-} - \gamma(\bar{\mathcal{B}}_{+} -
\bar{\mathcal{B}}_{-}) + \gamma^2\right]{\cal R}_-\,,\label{R1}\\
\Bigl(\xi{\frac d{d\xi}} - \bar{\mathcal{A}}_{+}\Bigr)\Bigl(\xi{\frac d{d\xi}} + \bar{\mathcal{A}}_{-}\Bigr)
{\cal R}_+ &= \left[-\,\bar{\mathcal{B}}_{+}\bar{\mathcal{B}}_{-} - \gamma(\bar{\mathcal{B}}_{+} -
\bar{\mathcal{B}}_{-}) + \gamma^2\right]{\cal R}_+\,.\label{R2}
\end{align}
Using the ansatz
\begin{equation}\label{Rv}
{\cal R}_\pm = \xi^{-1}\,v_{\mp{\frac 12}}\,, 
\end{equation}
we recast Eqs. (\ref{R1}) and (\ref{R2}) into
\begin{equation}\label{Dv}
\xi^2{\frac {d^2}{d\xi^2}}v_s + \Bigl[{\frac 14} - \bar{\mu}_f^2
- \beta^2\xi^2 - 2\beta\xi\,(\alpha + s)\Bigr]\,v_s = 0\,,
\end{equation}
where $s = \pm {\frac 12}$, as before, but now we have
\begin{align}
\bar{\mu}_f^2 &= \alpha^2 + \gamma^2 -\,\bar{\mathcal{B}}_{+}\bar{\mathcal{B}}_{-}
- \gamma(\bar{\mathcal{B}}_{+} - \bar{\mathcal{B}}_{-})\nonumber\\
&= \frac{1}{\mu^2}\Bigl[ -\,\omega^2\,{\frac {\kappa}{\sigma + \kappa}}
+ \left(k_3 + \Omega/2\right)^2 \Bigr]\,.\label{muf2}
\end{align}
With the independent variable $\tilde{\xi} = 2 |\beta| \xi$, Eq. (\ref{Dv}) can again be reduced to Whittaker's equation
\begin{equation}\label{Dv2}
\frac{d^2 v_s}{d\tilde{\xi}^2}  + \left[ -\,\frac{1}{4} + \frac{\tilde{\kappa}_f}{\tilde{\xi}}
+ \frac{\tfrac{1}{4} - \bar{\mu}_f^2}{\tilde{\xi}^2}\right] v_s = 0\,,
\end{equation}
where $\tilde{\kappa}_f$ is given by Eq.~\eqref{H29}. The regular solution of Eq.~\eqref{Dv2} is given by
\begin{equation}\label{vs}
v_s = v_s^0\, \exp(-\tfrac{1}{2} \tilde{\xi})\,\tilde{\xi}^{\tfrac{1}{2} + \bar{\mu}_f}
\,{}_1F_1(\tfrac{1}{2} + \bar{\mu}_f -\tilde{\kappa}_f, 1+2\bar{\mu}_f; \tilde{\xi})\,,   
\end{equation}
where  
\begin{equation}\label{kmn}
\frac{1}{2} + \bar{\mu}_f -\tilde{\kappa}_f = -\,n\,, \quad n = 0, 1, 2, \dots\,.
\end{equation}
As before, we can combine Eqs.~\eqref{kmn} and (\ref{muf2}) to derive the dispersion relation
\begin{equation}\label{disp2}
\omega = (2n + 1 + 2s)\Omega \pm \left[ (k_3 + \Omega/2)^2
- \frac{\kappa}{\sigma}(2n + 1 + 2s)^2 \Omega^2 \right]^{1/2}\,.
\end{equation}

The motion of Dirac waves in the G\"odel-type universe is in general agreement with the corresponding results for the scalar and electromagnetic wave propagation described in Appendices~\ref{appB} and \ref{appC}.

\subsection{Dealing with subtle points of Dirac theory on curved spacetimes}

In order to have a correct quantum-mechanical interpretation,  Dirac equation (\ref{H1}) should be recast into the form of the Schr\"odinger equation
\begin{equation}
i\frac{\partial \Psi} {\partial t}= {\cal H}\Psi\,.\label{sch}
\end{equation}
In flat spacetime with the Minkowski metric $\eta_{\mu\nu} = {\rm diag}(-1, 1, 1, 1)$ and the trivial frame $e^{\mu}{}_{\hat \alpha} = \delta^\mu_\alpha$ and spin connection $\Gamma_\mu = 0$, this is straightforward. Multiplying Eq.~\eqref{H1} by $\gamma^{\hat 0}$, we derive Schr\"odinger equation (\ref{sch}) with the Hermitian Hamiltonian 
\begin{eqnarray}
{\cal H} = \beta_{\rm D}\, m + \bm{\alpha}_{\rm D}\cdot\bm{p}\,.\label{hamM}
\end{eqnarray}
Here we denote, as usual, the matrices
\begin{equation}\label{beal}
\beta_{\rm D} := \gamma^{\hat 0}\,,\qquad \alpha_{\rm D}^{\hat i} := \gamma^{\hat 0}\gamma^{\hat i} = \begin{bmatrix}
0&\sigma_i \\ \sigma_i&0 \end{bmatrix}\,,\qquad i = 1,2,3\,.
\end{equation}
In addition, one also needs a quantum-probabilistic picture which is related to the normalization of the wave function. As is well-known, a direct consequence of the Dirac equation (\ref{H1}) is the conservation of the vector current, which in flat spacetime can be expressed as
\begin{equation}\label{dJ}
\partial_\mu J^\mu = 0\,,\qquad J^\mu = \overline{\Psi}\gamma^\mu\Psi\,. 
\end{equation}
Integration over 3-space yields a global conservation law
\begin{equation}\label{PP1}
\int d^3x\,J^0 = \int d^3x \Psi^\dagger\Psi = {\rm constant} = 1\,. 
\end{equation}
The physical interpretation of the Dirac fermion dynamics is based on Eqs.~\eqref{sch} and (\ref{PP1}), especially when the fermionic particle interacts with external fields. 

Dirac theory on curved manifolds, however, involves a number of subtleties. In particular, the differential conservation law (\ref{dJ}) is replaced by its curved version
\begin{equation}\label{dJ2}
\nabla_\mu J^\mu = {\frac {1}{\sqrt{-g}}}\,\partial_\mu\left(\sqrt{-g} J^\mu\right) = 0\,,
\qquad J^\mu = e^{\mu}{}_{\hat\alpha}\,\overline{\Psi}\gamma^{\hat\alpha}\Psi\,,
\end{equation}
which yields the global conservation law
\begin{equation}\label{PP2}
\int d^3x\,\sqrt{-g}\,J^0 = \int d^3x \sqrt{-g}\,e^{0}{}_{\hat\alpha}\Psi^\dagger\gamma^{\hat 0}\gamma^{\hat\alpha}\Psi = {\rm constant} = 1\,.
\end{equation}
For the natural G\"odel-type tetrad frame (\ref{G9}), we have $e^{0}{}_{\hat\alpha}\gamma^{\hat 0}\gamma^{\hat\alpha} = 1 -\sqrt{\frac{\sigma}{\sigma + \kappa}}\,\alpha_{\rm D}^{\hat 2}$; therefore, the physical interpretation of the solutions is unclear. Besides that, Dirac equation (\ref{DirL}) obviously cannot be directly recast into the form of the Schr\"odinger wave equation (\ref{sch}).

Both issues are related to the choice of the tetrad frame, which is defined up to an arbitrary local Lorentz transformation. The choice (\ref{G9}) corresponds to the so-called Landau-Lifshitz gauge with $e_{0}{}^{\hat i} = 0$ and $e^{i}{}_{\hat 0} = 0$. The situation is essentially improved when one chooses the Schwinger gauge for the frame, where $e_{i}{}^{\hat 0} = 0$ and $e^{0}{}_{\hat i} = 0$. Then, Eq.~\eqref{PP2} reduces to an ``almost flat'' form
\begin{equation}\label{PP3}
\int d^3x \sqrt{-g}\,e^{0}{}_{\hat 0}\,\Psi^\dagger\Psi = 1\,,
\end{equation}
and the Dirac equation is straightforwardly recast into the Schr\"odinger form \cite{Obukhov:2013zca,Obukhov:2017avp}. 

This suggests replacing the original tetrad frame (\ref{G9}) by a new one
\begin{align}
\widetilde{e}_{\hat 0} = \sqrt{\frac \kappa {\sigma + \kappa}}\Bigl(\partial_t + {\frac {\sqrt{\sigma}} \kappa}
e^{-\mu x}\partial_y\Bigr)\,,\quad \widetilde{e}_{\hat 1} =\partial_x\,,\quad \widetilde{e}_{\hat 2} =
{\frac {e^{-\mu x}}{\sqrt{\kappa}}}\partial_y\,,\quad \widetilde{e}_{\hat 3} = \partial_z,\label{fraS}
\end{align}
where we assume $\kappa > 0$. Obviously, this choice corresponds to the Schwinger gauge $\widetilde{e}_{i}{}^{\hat 0} = 0$ and $\widetilde{e}{\,}^{0}{}_{\hat i} = 0$, for $i = 1,2,3$.

For G\"odel-type spacetimes, the two frames (\ref{G9}) and (\ref{fraS})  are related by the Lorentz transformation,
\begin{equation}
\widetilde{e}{\,}_{\hat\alpha} = \Lambda^{\hat\beta}{}_{\hat\alpha}\,e_{\hat\beta},\label{SL}
\end{equation}
where explicitly
\begin{equation}
\Lambda^{\hat\alpha}{}_{\hat\beta} = \left(\begin{array}{cccc}
\sqrt{\frac {\sigma + \kappa} \kappa} & 0 & \sqrt{\frac \sigma \kappa} & 0 \\
0 & 1 & 0 & 0 \\ \sqrt{\frac \sigma \kappa} & 0 & \sqrt{\frac {\sigma + \kappa} \kappa} & 0 \\
0 & 0 & 0 & 1 \end{array}\right)\,.\label{Lor}
\end{equation}
Interestingly, the transformation with constant matrix elements is global, whereas in general only local Lorentz transformations are possible.

The change of a frame on the spacetime affects the fermionic wave function
\begin{equation}\label{PsiL}
\Psi\quad \longrightarrow\quad \widetilde{\Psi} = L^{-1}\,\Psi
\end{equation}
via the spinor matrix $L$ that satisfies
\begin{equation}\label{LL}
L^{-1}\gamma^{\hat\alpha}L = \Lambda^{\hat\alpha}{}_{\hat\beta}\,\gamma^{\hat\beta}.
\end{equation}
Using a convenient parametrization with $\cosh\zeta = \sqrt{\frac {\sigma + \kappa} \kappa}$ and $\sinh\zeta = \sqrt{\frac {\sigma} \kappa}$, we easily derive
\begin{equation}\label{traL}
L = \cosh(\zeta/2)\,I_4 + \sinh(\zeta/2)\,\alpha_{\rm D}^{\hat 2} =
\begin{bmatrix}
\cosh(\zeta/2)\,I_2 & \sinh(\zeta/2)\,\sigma_2 \\
\sinh(\zeta/2)\,\sigma_2 & \cosh(\zeta/2)\,I_2 
\end{bmatrix}
\,.
\end{equation}
The spinor transformation (\ref{PsiL}) mixes the spin-up and spin-down states (${\cal L}_\pm$) for the left modes (and similarly for the right modes) and an appropriate normalization of the solutions should be fixed for the squares $\widetilde{\Psi}^\dagger\widetilde{\Psi}$ of the transformed wave functions. 


\section{Dirac Equation in Fermi Frame}

Let us next consider the Dirac equation in the quasi-inertial Fermi frame of Section~\ref{FermiC}. We are interested in the propagation of Dirac particles as described by fundamental observers that are all spatially at rest in the Fermi frame and occupy the limited cylindrical region about the $Z$ axis such that $\Omega |X| = \Omega |Y| \lesssim \varepsilon$. As before, we ignore all terms of order $\varepsilon^4$ and higher. The preferred observers have adapted orthonormal tetrad frames $\varphi^{\mu}{}_{\hat \alpha}$ given in Eqs.~\eqref{F20}--\eqref{F21}. Let us note that  $\varphi_{\mu \,\hat \alpha}$ can be written in $(T, X, Y, Z)$ coordinate system as
\begin{equation}\label{J1}
\varphi_{\mu \, \hat 0} = ( -1 + \tfrac{1}{2}\hat{h}_{ 0 0}, \hat{h}_{01}, \hat{h}_{02}, 0)\,, \qquad \varphi_{\mu \, \hat 1} = ( 0, 1 + \tfrac{1}{2}\hat{h}_{1 1}, \hat{h}_{1 2}, 0)\,,        
\end{equation}
\begin{equation}\label{J2}
\varphi_{\mu \, \hat 2} = ( 0, 0, 1 + \tfrac{1}{2}\hat{h}_{2 2}, 0)\,, \qquad \varphi_{\mu \, \hat 3} = ( 0, 0, 0, 1)\,.
\end{equation}

We employ perturbations beyond Minkowski spacetime in our Fermi frame; hence, in the absence of $\hat{h}_{\mu \nu}$, we have $\varphi^{\mu}{}_{\hat \alpha} \to \delta^\mu _\alpha$. To simplify matters even further, we assume henceforth that the deviation from Minkowski spacetime is only due to  the gravitomagnetic potentials $\hat{h}_{0 1}= -\,{\frac{\kappa}{\sigma}}\,\Omega^3 Y(X^2 + Y^2)$ and $\hat{h}_{0 2}= {\frac{\kappa}{\sigma}}\,\Omega^3 X(X^2 + Y^2)$ that give rise to the gravitomagnetic field $\hat{\bm{B}} = (0, 0, \hat{B}_3)$, where $\hat{B}_3 = -\,2{\frac{\kappa}{\sigma}}\,\Omega^3 (X^2 + Y^2)$.

With these assumptions, the spin connection~\eqref{H7}  can be computed using the tetrad system $\varphi_{\mu \,\hat \alpha}$ that is adapted to our reference observers and we find 
\begin{equation}\label{J3}
\gamma^\mu \hat{\Gamma}_\mu = \frac{i}{2} \hat{B}_3 
\begin{bmatrix}
\sigma_3&0 \\
0&-\sigma_3 
\end{bmatrix}
\,. 
\end{equation}
That is, the spin connection is proportional to the gravitomagnetic field of the G\"odel-type universe in the Fermi frame under consideration here.  

For the sake of simplicity, we assume a solution of the Dirac equation that propagates along the $Z$ axis and is of the form
\begin{equation}\label{J4}
\hat{\Psi} = \hat{\psi}(X, Y) \exp (-i\,\omega\, T +i\, k_3\,Z)\,.
\end{equation}
Moreover, it is convenient to define 
\begin{equation}\label{J5}
\hat{\mathbb{X}} = \binom {\hat{\psi}_1} {\hat{\psi}_3}\,, \qquad \hat{\mathbb{Y}} = \binom {\hat{\psi}_2}{\hat{\psi}_4}\,.
\end{equation}
In this case, Dirac's equation reduces to 
\begin{equation}\label{J6}
\left[\partial_X + i \partial_Y + {\frac{\kappa}{\sigma}}\,\omega \Omega^3 (X^2+Y^2)(X+iY)\right] \hat{\mathbb{X}} = -\,{\frac{i\kappa}{\sigma}}\,\Omega^3 (X^2+Y^2)\sigma_1\hat{\mathbb{Y}} + i
\begin{bmatrix}
k_3&\omega+m \\
\omega - m&k_3 
\end{bmatrix}
\hat{\mathbb{Y}}\, 
\end{equation}
and
\begin{equation}\label{J7}
\left[\partial_X - i \partial_Y - {\frac{\kappa}{\sigma}}\,\omega \Omega^3 (X^2+Y^2)(X-iY)\right] \hat{\mathbb{Y}} = {\frac{i\kappa}{\sigma}}\,\Omega^3 (X^2+Y^2)\sigma_1\hat{\mathbb{X}} + i
\begin{bmatrix}
-k_3&\omega+m \\
\omega - m&-k_3 
\end{bmatrix}
\hat{\mathbb{X}}\,. 
\end{equation}
Here, $\partial_X := \partial /\partial X$, etc.; furthermore, we note that 
\begin{equation}\label{J8}
(\partial_X \pm i \partial_Y) (X^2+Y^2)^2 = 4(X^2+Y^2)(X \pm iY)\,,
\end{equation}
\begin{equation}\label{J9}
(\partial_X \pm i \partial_Y) [(X^2+Y^2)(X\mp iY)] = 4(X^2+Y^2)\,.
\end{equation}

In the absence of the gravitational perturbation, positive-frequency plane wave solutions of the free Dirac equation propagating in the $Z$ direction are given by
\begin{equation}\label{J10}
\hat{w}^{\pm}\, e^{-i\,\omega\, T +i\, k_3\,Z}\,,
\end{equation}
where the spin of the Dirac particle is either parallel ($\hat{w}^+$) or antiparallel ($\hat{w}^-$) to the $Z$ direction; that is,  
\begin{equation}\label{J11}
\hat{w}^{+} = N^{\Uparrow}
\begin{bmatrix}
1 \\
0 \\
\varrho  \\
0
\end{bmatrix}
\,, \qquad
\hat{w}^{-} = N^{\Downarrow}
\begin{bmatrix}
0 \\
1 \\
0  \\
-\varrho 
\end{bmatrix}
\,. 
\end{equation}
Here, $N^{\Uparrow}$ and $N^{\Downarrow}$ are positive normalization constants, $\omega = (m^2 + k_3^2)^{1/2}$ and 
\begin{equation}\label{J12}
\varrho := \frac{k_3}{\omega + m} = \frac{\omega - m}{k_3}\,.
\end{equation}

With these background states, we solve Eqs.~\eqref{J6} and~\eqref{J7} to linear order in the gravitomagnetic perturbation and obtain,  after some algebra,
\begin{equation}\label{J13}
\hat{\Psi}^{+} = N^{\Uparrow}
\begin{bmatrix}
\exp[-\,\tfrac{3\kappa}{8\sigma}\omega \Omega^3 (X^2+Y^2)^2] \\
\tfrac{i\kappa}{4\sigma} \Omega^3 \varrho (X^2+Y^2)(X+iY) \\
\varrho \exp[-\,\tfrac{3\kappa}{8\sigma}\omega \Omega^3 (X^2+Y^2)^2] \\
\tfrac{i\kappa}{4\sigma} \Omega^3 (X^2+Y^2)(X+iY)
\end{bmatrix}
e^{-i\,\omega\, T +i\, k_3\,Z}\,, 
\end{equation}
\begin{equation}\label{J14}
\hat{\Psi}^{-} = N^{\Downarrow}
\begin{bmatrix}
\tfrac{i\kappa}{4\sigma} \Omega^3 \varrho (X^2+Y^2)(X-iY) \\
\exp[\tfrac{3\kappa}{8\sigma}\omega \Omega^3 (X^2+Y^2)^2] \\
-\,\tfrac{i\kappa}{4\sigma} \Omega^3 (X^2+Y^2)(X-iY)  \\
-\varrho \exp[\tfrac{3\kappa}{8\sigma}\omega \Omega^3 (X^2+Y^2)^2]
\end{bmatrix}
e^{-i\,\omega\, T +i\, k_3\,Z}\,. 
\end{equation}
These solutions of Dirac's equation exhibit the coupling of spin with the gravitomagnetic field of G\"odel-type universe and may be compared and contrasted with the results of Appendix~\ref{appC} for the propagation of circularly polarized  electromagnetic waves along the $Z$ axis in the Fermi frame. 

We should note that fermions in G\"odel-type universes have been the subject of a number of previous studies; see, for instance,~\cite{Soares:1981, Leahy:1982, Soares:1985, Pimentel:1986, Villalba:1993, Pimentel:1994} and the references cited therein.

\section{Discussion}

Spin-gravity coupling represents a physically important subject matter in view of the basic nature of intrinsic spin of particles and the universality of the gravitational interaction. We have investigated in detail the coupling of intrinsic spin with the gravitomagnetic fields of a three-parameter class of G\"odel-type spacetimes.  These stationary and homogeneous rotating universes are characterized by the set of constant parameters $(\kappa, \sigma, \mu)$; for $\kappa < 0$, there are closed timelike curves (CTCs) in spacetime, while for $\kappa \ge 0$, CTCs are absent.  For $(\kappa, \sigma, \mu) \to (-1, 2, \sqrt{2}\, \Omega)$, we recover  G\"odel's rotating universe model, where $\Omega > 0$ is the frequency of rotation. On the background G\"odel-type spacetimes, we have studied Dirac's equation and worked out its solutions; furthermore, we have extended our results to exact Fermi normal coordinate systems in these universes.    We have shown that the Stern--Gerlach force due to the coupling of intrinsic spin with the gravitomagnetic field of a G\"odel-type spacetime is in agreement in the correspondence limit with the classical Mathisson spin-curvature force. This is a nonlinear generalization of previous work that focused on linearized general relativity~\cite{Mashhoon:2021qtc}. Our main results turn out to be independent of the possible causality difficulties of the G\"odel-type spacetimes.


\appendix


\section{Alternative Solution of Eq.~\eqref{H14}}\label{appA}

The purpose of this appendix is to present a different approach to the solution of Eq.~\eqref{H14}.

We can write Eq.~\eqref{H14} in the form
\begin{equation}\label{A1}
\xi\, \frac{d(\mathcal{U}\psi)}{d\xi} = \mathcal{U}\, \mathbb{M}\, \mathcal{U}^{-1} (\mathcal{U} \psi)\,, 
\end{equation}
where $\mathcal{U}$ is a constant unitary matrix given by
\begin{equation}\label{A2}
\mathcal{U} = \frac{1}{\sqrt{2}}
\begin{bmatrix}
I_2&- I_2 \\
I_2&I_2  
\end{bmatrix}
\,. 
\end{equation}
Under this similarity transformation, we have 
\begin{equation}\label{A3}
\mathcal{U}\, \gamma^{\hat 0}\, \mathcal{U}^{\dagger} = 
\begin{bmatrix}
0& I_2 \\
I_2&0  
\end{bmatrix}
= \gamma_5 \,, \qquad  \mathcal{U}\, \gamma^{\hat i}\, \mathcal{U}^{\dagger} = \gamma^{\hat i}\,.
\end{equation}
That is, the standard representation of Dirac matrices is thus transformed to the chiral (Weyl) representation. Employing this representation, we find
\begin{equation}\label{A4}
 \mathcal{U}\, \mathbb{M}\, \mathcal{U}^{-1} =
\begin{bmatrix}
-\bar{\mathcal{A}}_{+}&i\bar{\mathcal{B}}_{+}-i\gamma &0&0 \\
i\bar{\mathcal{B}}_{-}+i\gamma &\bar{\mathcal{A}}_{-}&0&0  \\
0&0&-\bar{\mathcal{A}}_{+}& -i\bar{\mathcal{B}}_{+}-i\gamma  \\
0&0&- i\bar{\mathcal{B}}_{-}+i\gamma &\bar{\mathcal{A}}_{-}
\end{bmatrix} - {\frac {im}{2\Omega}}\,\sqrt{\frac {\sigma}{\sigma + \kappa}}
\begin{bmatrix} 0&0&0&1 \\ 0&0&1&0  \\ 0&-1&0&0 \\ -1&0&0&0 \end{bmatrix}
\,, 
\end{equation}
where $\bar{\mathcal{A}}_{\pm}$ and $\bar{\mathcal{B}}_{\pm}$ are given by Eq.~\eqref{H16}. Expressing $\mathcal{U} \psi$ in the form
\begin{equation}\label{A5}
\mathcal{U} \psi= \sqrt{2}\,\begin{bmatrix} \cal R  \\ \cal L \end{bmatrix}, \qquad \cal R= \begin{bmatrix} \cal R_{+} \\ \cal R_{-} \end{bmatrix}, \qquad \cal L =\begin{bmatrix}
\cal L_{+} \\ \cal L_{-} \end{bmatrix}\,, 
\end{equation}
where $\cal R$ and $\cal L$ are now right-handed and left-handed two-component Weyl spinors, we recover system of equations~\eqref{l1}--\eqref{r2}. The rest of the analysis would follow the treatment presented in Section~\ref{Dirac}.  
 

\section{Scalar Waves in the G\"odel-type Universe}\label{appB}

Consider first a scalar field $\phi$ of inertial mass $m$ propagating on the background G\"odel-type spacetime (\ref{GT}). The wave equation is
\begin{equation}\label{B1}
g^{\mu \nu}  \phi_{; \mu \nu} - \frac{m^2c^2}{\hbar^2} \phi = 0\,,
\end{equation}
where $\hbar/(mc)$ is the Compton wavelength of the particle. The back reaction is of second order in the perturbation and can be neglected. The scalar wave equation can be written as 
\begin{equation}\label{B2}
\frac{1}{\sqrt{-g}}\,\frac{\partial}{\partial x^\mu} \left(\sqrt{-g} \,g^{\mu \nu}\frac{\partial \phi}{\partial x^\nu} \right)  - \frac{m^2c^2}{\hbar^2} \phi = 0\,,
\end{equation}
where for metric~\eqref{GT},  $\sqrt{-g} = e^{\mu x}\sqrt{\sigma + \kappa}$. Moreover, $\partial_t$, $\partial_y$ and $\partial_z$ are Killing vector fields; therefore, we assume
\begin{equation}\label{B3}
\phi(x) = e^{-i \omega t + i k_2 y +i k_3 z}\, \bar{\phi}(\xi)\,,\qquad \xi := e^{-\,\mu x}\,,
\end{equation}
where $\xi$ increases from $0$ to $\infty$ as the $x$ coordinate decreases from $+\infty$ to $-\infty$. In terms of the new radial variable $\xi$, the equation for $\bar{\phi}$ reduces to 
\begin{equation}\label{B4}
\frac{d^2 \bar{\phi}}{d\xi^2}  - \left[ \alpha_s^2 +\frac{\beta_s}{\xi} +
\frac{\zeta_s(\zeta_s + 1)}{\xi^2}\right] \bar{\phi} = 0\,,
\end{equation}
where 
\begin{equation}\label{B5}
\alpha_s = \frac{k_2}{\mu\sqrt{\sigma + \kappa}}\,, \quad \beta_s = \frac{2\omega\,k_2}{c\,\mu^2}\,
{\frac {\sqrt{\sigma}}{\sigma + \kappa}}, \quad \zeta_s(\zeta_s + 1) = \frac{1}{\mu^2}\left(
-\,{\frac {\omega^2}{c^2}}\,{\frac {\kappa}{\sigma + \kappa}} + k_3^2 + {\frac {m^2c^2}{\hbar^2}}\right).
\end{equation}

Let us assume $\zeta_s > 0$ and note that for $k_2 = 0$, Eq.~\eqref{B4} for $\bar{\phi}$ has solutions of the form $\xi^{-\zeta_s}$ and $\xi^{\zeta_s +1}$ that diverge at $\xi = 0$ and $\xi = \infty$, respectively. However, the scalar perturbation must be finite everywhere; therefore, waves cannot freely propagate parallel or antiparallel to the axis of rotation of the G\"odel-type spacetime. Next, for $k_2 \ne 0$, we introduce a new variable $\bar{\xi} := {\frac {|k_2|\,\sqrt{\sigma}}{\Omega(\sigma + \kappa)}}\,\xi$, in terms of which Eq.~\eqref{B4} takes the form of Whittaker's equation~\cite{A+S},
\begin{equation}\label{B6}
 \frac{d^2 \bar{\phi}}{d\bar{\xi}^2} +\left[ -\,\frac{1}{4}  +\frac{\bar{\kappa}_s}{\bar{\xi}} + \frac{\tfrac{1}{4} -\bar{\mu}_s^2}{\bar{\xi}^2}\right] \bar{\phi} = 0\,,
\end{equation}
where 
\begin{equation}\label{B7}
\bar{\kappa}_s = -\,\frac{\omega}{2\Omega}\,\frac{k_2}{|k_2|}\,{\frac {\sigma}{\sigma + \kappa}}\,,
\qquad \bar{\mu}_s = \pm (\zeta_s + \tfrac{1}{2})\,.
\end{equation}
In terms of the confluent hypergeometric functions, bounded solutions of this equation can be expressed up to proportionality constants by 
\begin{equation}\label{B8}
\exp(-\,\tfrac{1}{2} \bar{\xi})\, \bar{\xi}^{\zeta_s + 1}\, _1F_1(-n, 2\zeta_s + 2; \bar{\xi})\,, \qquad n = 0, 1, 2, \dots\,.
\end{equation}
Here, $\zeta_s > 0$, $\bar{\mu}_s = \zeta_s + 1/2$ and
\begin{equation}\label{B9}
\zeta_s + 1 + \frac{\omega}{2\Omega}\,\frac{k_2}{|k_2|}\,{\frac {\sigma}{\sigma + \kappa}} = -\,n\,,
\qquad  \omega = \pm \,2\Omega\,( n + \zeta_s + 1)\,{\frac {\sigma + \kappa}{\sigma}}\,,
\end{equation} 
for $k_2 < 0$ (upper plus sign) or $k_2 > 0$ (lower minus sign), respectively. Negative frequency in the case of $k_2 > 0$ indicates that waves traveling forward in time move backward along the $y$ direction. Finally, we note that only certain frequencies are allowed for the scalar waves; for instance, for   $k_2 < 0$ , we have $\omega_n = 2\Omega\,( n + \zeta_s + 1)(\sigma + \kappa)/\sigma$. That is, 
\begin{equation}\label{B10}
\omega_n^{\pm} = (2n + 1)\Omega \pm \left[\Bigl(-4n(n+1)\,{\frac {\kappa}\sigma} + 1\Bigr)\Omega^2
+ k_3^2 + \frac{m^2c^2}{\hbar^2}\right]^{1/2}\,,
\end{equation} 
where $\omega_n^+ > 0$ for all $n$ by definition, while  $\omega_n^- > 0$ for $n = 1, 2, 3, \dots$,  only if
\begin{equation}\label{B11}
\omega_n^+ \,\omega_n^- = 4n(n + 1)\,{\frac {\sigma + \kappa}\sigma}\,\Omega^2
-  k_3^2 - \frac{m^2c^2}{\hbar^2} > 0\,.
\end{equation}  

For further work on the scalar perturbations of the G\"odel-type universe and its extensions, see~\cite{Hisc, Thak, Saib1}.


\section{Electromagnetic Waves in the G\"odel-type Universe}\label{appC}

Propagation of electromagnetic radiation in the G\"odel universe was originally investigated in the search for the coupling of photon helicity with the rotation of matter~\cite{MashB}. In G\"odel-type spacetimes, Maxwell's equations can be reduced to an equation of the form of Eq.~\eqref{B4}, where instead of the quantities in Eq.~\eqref{B5}, we find \cite{KoOb2}
\begin{equation}\label{C1}
\alpha_s \to \alpha_{em} = \frac{K^{\pm}_2}{\mu\sqrt{\sigma + \kappa}}\,, \qquad \beta_s \to \beta_{em}
= \frac{2\omega\,K^{\pm}_2}{c\,\mu^2}\,{\frac {\sqrt{\sigma}}{\sigma + \kappa}}\,
\end{equation}
and $\zeta_s \to \zeta_{em}$, where
\begin{equation}\label{C2}
\zeta_{em}(\zeta_{em} + 1) = \frac{1}{\mu^2}\left(-\,{\frac {\omega^2}{c^2}}\,{\frac {\kappa}{\sigma + \kappa}}
+ (K^{\pm}_3)^2 \mp 2\Omega\,K^{\pm}_3\right)\,,
\end{equation}
since photon is massless ($m = 0$). The helicity coupling evident in Eq.~\eqref{C2} is consistent with  the spin-vorticity-gravity coupling described in Section~\ref{SVG}. That is, based on the results of Section~\ref{SVG}, we would expect that the corresponding Hamiltonian for a photon to be proportional to $\pm\, \hbar K^{\pm}_3 \Omega/\omega$, so that in terms of frequency we would have $\pm\, K^{\pm}_3 \Omega/\omega$. The effect should disappear in the case of a null geodesic consistent with the eikonal limit $\omega \to \infty$.  For further extensions and generalizations to G\"odel-type universes, see~\cite{CVD, ObKo, KoOb, KoOb2, Abd:1993, Havare:2002, Saib2}. 

\subsection{EM Waves in the Fermi Frame}

We consider the propagation of electromagnetic radiation on the background quasi-inertial Fermi normal coordinate system. In terms of the Faraday tensor $F_{\mu \nu}$, the source-free Maxwell equations can be expressed as
\begin{equation}\label{C3}
F_{[\mu \nu , \rho]} = 0\,, \qquad  (\sqrt{-g}\,F^{\mu \nu})_{,\nu} = 0\,. 
\end{equation}
Using the same approach as in~\cite{MashB}, we replace the gravitational field by a hypothetical optical medium that occupies Euclidean space with Cartesian Fermi coordinates $(X, Y, Z)$. The electromagnetic field equations~\eqref{C3} reduce to the  traditional form of Maxwell's equations in a medium with the decompositions 
\begin{equation}\label{C4}
F_{\mu \nu} \to (\tilde{\bm{E}}, \tilde{\bm{B}})\,, \qquad \sqrt{-g}\,F^{\mu \nu} \to (-\tilde{\bm{D}}, \tilde{\bm{H}})\,. 
\end{equation}
That is,  $F_{0i} = -\tilde{E}_i$ and $F_{ij} = \epsilon_{ijk}\,\tilde{B}_k$; similarly, $\sqrt{-g}\,F^{0i} = \tilde{D}_i$ and $\sqrt{-g}\,F^{ij} = \epsilon_{ijk}\,\tilde{H}_k$.  Here, $\epsilon_{ijk}$ is the  totally antisymmetric symbol with $\epsilon_{123} = 1$. The corresponding optical medium turns out to be gyrotropic with constitutive relations~\cite{Sk, Pl, Fe, VIS, HeOb}
\begin{equation}\label{C5}
\tilde{D}_i = \hat{\epsilon}_{ij}\,\tilde{E}_j - (\hat{\bm{G}} \times \tilde{\bm{H}})_i\,, \qquad \tilde{B}_i = \hat{\mu}_{ij}\,\tilde{H}_j + (\hat{\bm{G}} \times \tilde{\bm{E}})_i\,,
\end{equation} 
where the characteristics of the medium are conformally invariant and are given by
\begin{equation}\label{C6}
\hat{\epsilon}_{ij} = \hat{\mu}_{ij} = -\sqrt{-\hat{g}}\,\frac{\hat{g}^{ij}}{\hat{g}_{00}}\,, \qquad \hat{G}_i = - \frac{\hat{g}_{0i}}{\hat{g}_{00}}\,.
\end{equation} 

Expressing electromagnetic fields in the standard complex form and introducing the Riemann--Silberstein vectors,
\begin{equation}\label{C7}
\tilde{\bm{F}}^{\pm} = \tilde{\bm{E}} \pm i\,\tilde{\bm{H}}\,, \qquad  \tilde{\bm{S}}^{\pm} = \tilde{\bm{D}} \pm i\,\tilde{\bm{B}}\,,
\end{equation} 
the wave propagation equation can be expressed as the Dirac equation for photons in the gravitational field. That is, 
\begin{equation}\label{C8}
\bm{\nabla} \times \tilde{\bm{F}}^{\pm} = \pm\, i\,\frac{\partial \tilde{\bm{S}}^{\pm}}{\partial t}\,, \qquad \bm{\nabla} \cdot \tilde{\bm{S}}^{\pm} = 0\,, 
\end{equation} 
where 
\begin{equation}\label{C9}
\tilde{S}^{\pm}_p = \hat{\epsilon}_{pq}\,\tilde{F}^{\pm}_q  \pm i\, (\hat{\bm{G}} \times \tilde{\bm{F}}^{\pm})_p\,. 
\end{equation}
The Dirac-type equation implies $\partial_t(\bm{\nabla} \cdot \tilde{\bm{S}}^{\pm}) = 0$; therefore, if $\bm{\nabla}\cdot \tilde{\bm{S}}^{\pm} = 0$ initially, then it is valid for all time.

To interpret the physical meaning of these results, it proves useful to consider plane electromagnetic waves of frequency $\omega$ propagating along the $z$ axis in a global inertial frame with coordinates $x^\mu = (t, \bm{x})$ in Minkowski spacetime. Maxwell's equations are linear; therefore, we can use complex electric and magnetic fields and use the convention that only the real parts correspond to measurable quantities. The waves can have two independent orthogonal linear polarization states along the $\hat{\bm{x}}$ and $\hat{\bm{y}}$ directions, where $\hat{\bm{x}}$ is a unit vector along the $x$ axis, etc. The circular polarization states are constructed from the linear polarization states via superposition; in this case, the electric ($\bm{e}$) and magnetic ($\bm{b}$) fields can be expressed as
\begin{equation}\label{C10}
\bm{e}_{\pm} = \frac{1}{2}a_{\pm}\,(\hat{\bm{x}} \pm i\, \hat{\bm{y}})\,e^{-i\omega\,(t-z)}\,, \qquad
\bm{b}_{\pm} = \mp\, \frac{i}{2}\, a_{\pm}\,(\hat{\bm{x}} \pm i\, \hat{\bm{y}})\,e^{-i\omega\,(t-z)}\,,
\end{equation} 
where $a_{+}$ and $a_{-}$ are constant complex amplitudes. Here, the upper (lower) sign represents waves in which the orthogonal electric and magnetic fields rotate in the positive (negative) sense about the direction of wave motion. In the case of a photon with positive (negative) circular polarization, the photon has positive (negative) helicity, namely,  its spin is $+ \hbar$ ($- \hbar$) along its direction of propagation. The Riemann--Silberstein vectors have interesting behaviors for helicity states of the photon; in fact, for \emph{positive-helicity} radiation,  
\begin{equation}\label{C11}
\bm{e}_{+} + i\, \bm{b}_{+} = a_{+}\,(\hat{\bm{x}} + i\, \hat{\bm{y}})\,e^{-i\omega\,(t-z)}\,, \qquad \bm{e}_{+} - i\, \bm{b}_{+}  = 0\,,
\end{equation} 
while for  radiation with \emph{negative helicity}, 
\begin{equation}\label{C12}
\bm{e}_{-} + i\, \bm{b}_{-}  = 0\,, \qquad \bm{e}_{-} - i\, \bm{b}_{-} = a_{-}\,(\hat{\bm{x}} - i\, \hat{\bm{y}})\,e^{-i\omega\,(t-z)}\,.
\end{equation} 
Hence, $\bm{e} + i\, \bm{b}$ ($\bm{e} - i\, \bm{b}$) represents in essence an electromagnetic wave with positive (negative) helicity. It is important to note that  Eqs.~\eqref{C8} and~\eqref{C9} that represent  the propagation of electromagnetic test fields in a gravitational field completely decouple for different helicity states. 

Imagine the propagation of electromagnetic waves with definite helicity along the $Z$ axis in the Fermi normal coordinate system in the G\"odel-type spacetime. The universe rotates in the negative sense about the $Z$ axis. We confine our considerations to the cylindrical region near the rotation axis where the perturbation analysis contained in Eqs.~\eqref{F17}--\eqref{F19} is valid. To simplify matters, we take into account only the gravitomagnetic potentials $\hat{h}_{01}$ and $\hat{h}_{02}$ and ignore the other potentials; therefore, in Eq.~\eqref{C6} we have 
\begin{equation}\label{C13}
\hat{\epsilon}_{ij} = \hat{\mu}_{ij} \approx 1\,, \qquad  \hat{\bm{G}} \approx -\,\frac{\kappa}{\sigma} \Omega^3 (X^2+Y^2) (Y, -X, 0)\,.
\end{equation} 
It is straightforward to show that in this case the field Eqs.~\eqref{C8} and~\eqref{C9} have the solution
\begin{equation}\label{C14}
\tilde{F}_1^{\pm} =  \hat{a}_{\pm} \exp[ -i\omega(T-Z) \mp \frac{\kappa}{4\sigma}\omega \Omega^3\,(X^2+Y^2)^2]\,,
\end{equation} 
\begin{equation}\label{C15}
\tilde{F}_2^{\pm} = \pm i \hat{a}_{\pm} \exp[ -i\omega(T-Z) \mp \frac{\kappa}{4\sigma}\omega \Omega^3\,(X^2+Y^2)^2]\,
\end{equation}
and $\tilde{F}_3^{\pm} = 0$. Here, $\hat{a}_+$ and $\hat{a}_-$ are constant amplitudes for the positive and negative helicity waves in the Fermi frame, respectively. If the wave propagates along the axis of rotation (i.e., $-Z$ direction), then in Eqs.~\eqref{C14} and~\eqref{C15} we have $Z \to -Z$ and $\pm \to \mp$ in the exponents of these equations as well as in the coefficient of the latter equation. For $\Omega = 0$, the Fermi frame reduces to a global inertial frame in Minkowski spacetime and we recover waves of the form given in Eqs.~\eqref{C10}--\eqref{C12}. 

The helicity-gravitomagnetic field coupling is evident in these results and corresponds to Eqs.~\eqref{F23} and ~\eqref{F24} of Section~\ref{FermiC}; indeed, the form of this coupling is reminiscent of the helicity-twist coupling  studied in~\cite{Bini:2018iyu}.


\end{document}